\documentclass[12pt,a4paper]{article}
\usepackage[british]{babel}
\usepackage{epsfig}
\usepackage{amsfonts, amssymb, latexsym, amsmath}
\usepackage{color}
\usepackage{dsfont}
\newcommand{\be}{\begin{equation}}
\newcommand{\ee}{\end{equation}}
\newcommand{\half}{\frac{1}{2}}

\newcommand{\Wpm}{W_\pm}
\newcommand{\Qpm}{Q_\pm}
\newcommand{\Qp}{Q_{+}}
\newcommand{\Qm}{Q_{-}}
\newcommand{\Wp}{W_{+}}
\newcommand{\Wm}{W_{-}}
\begin{document}
\date{}

\begin{titlepage}

\title{
{\vspace*{-6mm} \normalsize
\hfill \parbox{40mm}{DESY 04-098     \\
                     Bicocca-FT-04-8 \\
                     SFB/CPP-04-18   \\
                     MS-TP-04-13\\[25mm] }}\\[-12mm]
Twisted mass quarks and the phase structure of \\ lattice QCD
\vspace*{5mm}}

\author{F.\ Farchioni$^{a}$,
        R.\ Frezzotti$^{b}$,
        K.\ Jansen$^{c}$,
        I.\ Montvay$^{d}$,
        G.C.\ Rossi$^{c,e}$,\\
        E.\ Scholz$^{d}$,
        A.\ Shindler$^{c}$,
        N.\ Ukita$^{d}$,
        C.\ Urbach$^{c,f}$,
        I.\ Wetzorke$^{c}$ \\[5mm]
  {\small $^a$ Universit\"at M\"unster,
   Institut f\"ur Theoretische Physik,}\\
   {\small Wilhelm-Klemm-Strasse 9, D-48149 M\"unster,
   Germany}\\
  {\small $^b$ INFN, Sezione di Milano and 
   Dipartimento di Fisica, Universit\`a di Milano ``{\it Bicocca}''}\\
   {\small Piazza della Scienza 3 - 20126 Milano, Italy}\\
  {\small $^c$ NIC/DESY Zeuthen, Platanenallee 6, D-15738 Zeuthen,
   Germany}\\
  {\small $^d$ Deutsches Elektronen-Synchrotron DESY, Notkestr.\,85,
   D-22603 Hamburg, Germany}\\
  {\small $^e$ Dipartimento di Fisica, Universit\`a di  Roma
   ``{\it Tor Vergata}'' and INFN, Sezione di Roma 2}\\
  {\small Via della Ricerca Scientifica - 00133 Roma, Italy}\\
  {\small $^f$ Freie Universit\"at Berlin,
   Institut f\"ur Theoretische Physik,}\\
  {\small Arnimallee 14, D-14196 Berlin, Germany}}
%
%%%%%%%%%%%%%%%%%%%%%%%%%%%%%%%%%%%%%%%%%%%%%%%%%%%%%%%%%%%%%%%%%%%%%%%%

\maketitle
\vspace*{7mm}
\abstract{The phase structure of zero temperature twisted mass lattice
 QCD is investigated.
 We find strong metastabilities in the plaquette observable when the
 untwisted quark mass assumes positive or negative values.
 We provide interpretations of this phenomenon in terms of chiral
 symmetry breaking and the effective potential model of Sharpe and
 Singleton.}
\end{titlepage}

\newpage\setcounter{page}{2}
%%%%%%%%%%%%%%%%%%%%%%%%%%%%%%%%%%%%%%%%%%%%%%%%%%%%%%%%%%%%%%%%%%%%%%%%
\section{Introduction}\label{sec1}

 As a consequence of (soft) chiral symmetry breaking, nature has
 arranged itself such that three of the pseudo-scalar mesons are light,
 with masses around 140 MeV.
 This lightness of the pion mass becomes important also  when we think
 of numerical simulations in lattice QCD.
 Approaching the ``physical point'', at which the pion mass assumes its
 value as measured in experiment, the algorithms used in lattice
 simulations suffer from a substantial slowing down
 \cite{BERLIN,KARLTSUKUBA} which restricts present simulations to rather
 high and unphysical values of the quark mass.

 In addition to this slowing-down of the algorithms for Wilson fermions,
 the quark mass does not act as an infrared regulator allowing thus for
 the appearance of very small unphysical eigenvalues of the lattice
 Wilson-Dirac operator.
 These eigenvalues render the simulations more difficult and sometimes
 even impossible.

 Staggered fermions solve this problem but it is not clear how to use
 this approach to simulate $N_f=2$ or odd number of flavours
 \cite{KNECHTLI}.
 Overlap fermions \cite{OVERLAP} also solve the problem but they are
 computationally very demanding and, unless new algorithms are invented,
 they are very difficult to use for dynamical simulations.

 An elegant way out may be the use of so-called twisted mass fermions
 \cite{TMQCD,ALPHA-TMQCD}.
 This formulation of lattice QCD is obtained when the Wilson term and
 the physical quark mass term are taken not parallel in flavour chiral
 space, but rotated by a relative twist angle $\omega$.
 If the Wilson term is given the usual form, such a chiral rotation
 leads to a {\em twisted mass} parameter $\mu$, in addition to the
 standard Wilson quark mass $m_0$ (``untwisted'' quark mass).
 Lattice QCD with a twisted mass was first employed for
 ${\cal O}(a)$-improved Wilson fermions with the nice feature that the
 improvement coefficients and the renormalisation constants are the same
 as for ${\cal O}(a)$-improved Wilson fermions without twisted mass and
 hence they did not need to be recalculated \cite{TMQCD-ORDERA}.
 The main advantage of the twisted mass fermions is that the twisted
 quark mass provides a natural infrared cut-off and avoids problems
 with accidental small eigenvalues rendering therefore the simulations
 safe.
 Of course, the slowing down of the algorithms when approaching
 small quark masses will remain, although it is expected to be less
 severe.

 Later on it was realized that a full ${\cal O}(a)$-improvement of
 correlation functions can be obtained by using the twisted mass alone
 {\em without additional improvement terms} when, as a special case,
 $m_0$ is set to the critical value $m_{\mathrm{crit}}$ and the above
 mentioned twist angle is equal to $\omega=\pi/2$ \cite{FREZZOTTI-ROSSI}.
 In this way the demanding computation of many improvement coefficients
 can be avoided rendering the simulations much easier both from a
 conceptual as well as from a practical point of view.

 The Wilson twisted mass formulation has been tested numerically in the
 quenched approximation already \cite{SCALING}.
 The results are very encouraging.
 The ${\cal O}(a)$ corrections appear indeed to be cancelled and even
 higher order effects seem to be small, at least for the quantities and
 the value of the quark mass considered in ref.~\cite{SCALING}.

 A word of caution has to be added at this point.
 Although, as mentioned above, the twisted quark mass can be 
 decreased towards zero without simulations breaking down due to exceptional
 configurations, there is an important interplay between the lattice
 cut-off, $\Lambda = a^{-1}$ with $a$ the lattice spacing, and the quark
 mass $m_q$ (see equation (\ref{phtotm}) below).
 In the continuum in the presence of spontaneous chiral symmetry
 breaking the chiral symmetry is not realized {\it \`a la}
 Wigner and, as the quark mass goes to zero, the chiral phase of the
 vacuum is driven by the phase of the quark mass term.
 The same must be true on the lattice, thus the scaling limit $a\to 0$
 should be taken before letting $m_q\to 0$.
 As a result, taking the chiral limit is a numerically delicate matter.

 In order to ensure in practice that on the lattice the chiral phase of
 the vacuum is determined by the quark mass term, proportional to $m_q$,
 and not by the Wilson term, the lattice parameters should satisfy
 the order of magnitude inequality \cite{FREZZOTTI-ROSSI}
\begin{equation}\label{COND1}
m_q \Lambda_{\rm QCD}^{-1} \, \gg \, a \Lambda_{\rm QCD} \, .
\end{equation}
 This same condition emerges from many different corners of the lattice
 theory when the physical world is approached.
 A very simple argument leading to the bound~(\ref{COND1}) is
 obtained by comparing the magnitude of the critical Wilson
 term to that of the quark mass term and requiring the first to be
 negligibly small compared to the second one, in order to be sure
 that lattice physics matches the requirements of the continuum theory.
 From the order of magnitude inequality
 $a(\Lambda_{\rm QCD})^{5} \, \ll \,m_q (\Lambda_{\rm QCD})^3$,
 one immediately gets the condition~(\ref{COND1}).
 It is important to observe, however, that the less restrictive
 condition
\begin{equation}\label{COND2}
m_q \Lambda_{\rm QCD}^{-1} \, \gg \, (a \Lambda_{\rm QCD})^2 
\end{equation}
 may be sufficient if one is dealing with ${\cal O}(a)$-improved
 quantities.

 It should be remarked that, since $a\Lambda_{\rm QCD}$  
 can be (non-perturbatively) expressed in terms 
 of $g_0^2$, eqs.~(\ref{COND1}) and~(\ref{COND2}) are actually 
 (order of magnitude) conditions for the values of the dimensionless
 bare lattice parameters $am_q$ and $g_0^2$. Contact with dimensionful 
 quantities can be made by comparing simulation data with physical inputs.

 What is in practice important is 
 to know for which range of the bare lattice parameters 
 one can avoid troubles from chiral breaking cutoff effects, even if 
 parametrically of order $a^2$ or higher. 
 This issue has to be settled by numerical investigations aimed at
 establishing both the structure of the phase diagram of the lattice 
 model in study and the size of residual scaling violations on the 
 physical  observables. 
 
 In this perspective, twisted mass fermions offer a unique opportunity to 
 explore the phase diagram of Wilson fermions.
 By fixing the twisted mass parameter $\mu$, one may vary
 $(m_0 - m_{\mathrm{crit}})$ from positive to negative values.
 In this way, the phase diagram of zero temperature lattice QCD can be
 explored. 
 It should be emphasized that, on large lattices, such an investigation
 would be very difficult without having $\mu \ne 0$, since else the
 algorithms would slow down dramatically approaching the critical quark
 mass.

 In this work we have performed simulations to explore the phase diagram
 of zero temperature QCD. 
 As it will be shown in the following, we find strong metastabilities
 in the plaquette expectation value.
 We determined in both metastable branches a number of quantities such
 as the (untwisted) PCAC quark mass and pseudo-scalar meson masses.
 The results presented in this paper are obtained at only one value of
 $\beta=5.2$, with $\beta$ related to the bare gauge coupling $g_0$ by
 $\beta=6/g_0^2$.
 Since the value of $\beta=5.2$ corresponds to a rather coarse value of
 the lattice spacing ($a\approx 0.16\, {\rm fm}$) our work can only be
 considered as a starting point for a more detailed investigation of the
 phase diagram of lattice QCD.
 In particular, the $\beta$-dependence of the strength of the observed
 metastabilities has to be determined.
 We believe that a qualitative and even quantitative understanding of
 the phase diagram is a necessary prerequisite for phenomenologically
 relevant numerical simulations.

 The paper is organized as follows.
 In section \ref{sec2} we introduce Wilson twisted mass fermions and
 give our notations.
 This is followed by a short discussion of the algorithms used.
 In section \ref{sec3} we provide our evidence for metastabilities by
 hysteresis effects and long living metastable states.
 There, we also show results for a selected set of physical quantities.
 In section \ref{sec4}, we give a possible interpretation of these
 results in terms of chiral symmetry breaking and the Sharpe-Singleton
 effective potential model.
 We conclude finally in section~\ref{sec5}.
 In the appendix some details of the applied update algorithms are
 explained.

%%%%%%%%%%%%%%%%%%%%%%%%%%%%%%%%%%%%%%%%%%%%%%%%%%%%%%%%%%%%%%%%%%%%%%%%
\section{Lattice action and basic variables}\label{sec2}

%%%%%%%%%%%%%%%%%%%%%%%%%%%%%%%%%%%%%%%%%%%%%%%%%%%%%%%%%%%%%%%%%%%%%%%%
\subsection{Lattice action}\label{sec2.1}

 Let us start by writing the Wilson tmQCD action as
\begin{equation}
  \label{tmaction}
  S[U,\chi,\bar\chi] = \bar\chi\left(D[U] + m_0 + \mu i 
\gamma_5\tau_3\right)\chi\; .
\end{equation}
 In eq.~(\ref{tmaction}) $m_0$ is the {\em quark mass} parameter
 and $\mu$ is the {\em twisted quark mass} parameter.
 The operator $D[U]$ is given by
\begin{eqnarray}
\label{DofU}
\bar{\chi}\, D[U]\, \chi 
& = & a^4\sum_x \left\{\frac{4r}{a}\bar\chi(x)\chi(x)\right.\\
& - & \left. \frac{1}{2a}\bar\chi(x) \sum_{\mu=1}^4 \Big(
                   U(x,\mu)(r + \gamma_\mu) \chi(x+a\hat\mu) +
                U^\dagger(x-a\hat\mu,\mu)(r-\gamma_\mu)\chi(x-a\hat\mu)
                \Big)\right\}\; , \nonumber
\end{eqnarray}
 with $r$ the Wilson parameter which will be set to $r=1$ in our
 simulations.

 The action as it stands in eq.~(\ref{tmaction}) can, of course, be
 studied in the full parameter space $(m_0,\mu)$.
 A special case arises, however, when $m_0$ is tuned towards a critical
 bare quark mass $m_{\mathrm{crit}}$.
 In such, and only in such a situation all physical quantities are, or
 can easily be, $\mathcal{O}(a)$ improved.
 It is hence natural to rewrite
\be
\label{m0}
m_0=m_{\mathrm{crit}} + \tilde{m}
\ee
 with $\tilde{m}$ an offset quark mass.
 The values of $m_{\mathrm{crit}}$ need only to be known with
 $\mathcal{O}(a)$ accuracy \cite{FREZZOTTI-ROSSI} and can be, for
 instance, taken from the pure Wilson theory at $\mu=0$.

 For standard Wilson fermions usually the hopping parameter
 representation is taken in the numerical simulations.
 This representation is easily obtained from eq.~(\ref{tmaction}) by a
 rescaling of the fields
\be
\label{kappatrafo}
\chi \to \frac{\sqrt{2\kappa}}{a^{3/2}} \chi , \hspace*{3em}
\bar\chi \to \frac{\sqrt{2\kappa}}{a^{3/2}} \bar\chi , \hspace*{3em}
\kappa = \frac{1}{2am_0 + 8r}\; .
\ee
 We then obtain the form of the action that is actually used in our
 simulations
\begin{eqnarray}
\label{eq2.1:6}
S[\chi,\bar\chi,U] & = & \sum_x
\left\{\bar\chi(x)\Big( 1 + 2ia\mu\kappa \gamma_5 \tau_3 \Big) 
\chi(x) \right. \\
& - & \left. \kappa\bar\chi(x)
\sum_{\mu=1}^4 \Big(U(x,\mu)(r + \gamma_\mu)\chi(x+a\hat\mu) +
U^\dagger(x-a\hat\mu,\mu)(r-\gamma_\mu)\chi(x-a\hat\mu)\Big)
\right\} \ . \nonumber
\end{eqnarray}

 Although not needed for the discussion of the numerical data presented
 below, we give for completeness here the action in the so-called
 {\em physical basis}.
 This action is obtained by introducing new fields $\psi(x)$ and
 $\bar{\psi}(x)$ which are related to the fields in
 eq.~(\ref{tmaction}) by a chiral transformation
\begin{eqnarray}\label{transformation}
\psi(x) &\equiv& e^{i\frac{\omega}{2}\gamma_5\tau_3}\chi(x) =
\left( \cos\frac{\omega}{2} +
i\gamma_5 \tau_3 \sin\frac{\omega}{2} \right) \chi(x) \ ,
\nonumber \\[0.5em]
\bar{\psi}(x) &\equiv&
\bar{\chi}(x) e^{i\frac{\omega}{2}\gamma_5\tau_3} =
\bar{\chi}(x) \left( \cos\frac{\omega}{2} +
i\gamma_5 \tau_3 \sin\frac{\omega}{2} \right) \ .
\end{eqnarray}
The action then reads
\begin{eqnarray}\label{phaction}
S[\psi,\bar\psi,U] & = & a^4\sum_x
\left\{ m_q \bar\psi(x) \psi(x) -
\frac{1}{2a}\bar\psi(x){\rm{e}}^{-i\omega\gamma_5\tau_3} 
\times \right. \nonumber \\
& &
\left. \Big[ \sum_{\mu=1}^4 \Big(rU(x,\mu)\psi(x+a\hat\mu) +
rU^\dagger(x-a\hat\mu,\mu)\psi(x-a\hat\mu)\Big) -
(2am_{\mathrm{crit}}+8r)\psi(x)\Big]  \right. \nonumber \\
& &
\left. - \frac{1}{2a}\bar\psi(x)
\sum_{\mu=1}^4 \Big(U(x,\mu)\gamma_\mu\psi(x+a\hat\mu) -
U^\dagger(x-a\hat\mu,\mu)\gamma_\mu\psi(x-a\hat\mu)\Big)
\right\}
\end{eqnarray}
 where we have identified
\be\label{phtotm}
m_q\cos\omega  =  m_0 - m_{\mathrm{crit}} = \tilde{m}, \hspace*{3em}
m_q\sin\omega = \mu \ .
\ee

%%%%%%%%%%%%%%%%%%%%%%%%%%%%%%%%%%%%%%%%%%%%%%%%%%%%%%%%%%%%%%%%%%%%%%%%
\subsection{Simulation algorithms}\label{sec2.2}

 In our numerical simulations we used two different optimized updating
 algorithms for producing samples of gauge configurations:
 the Hybrid Monte Carlo (HMC) algorithm with up to three pseudo-fermion
 fields as suggested in \cite{HASENBUSCH,HASENBUSCH-JANSEN} and the
 two-step multi-boson (TSMB) algorithm \cite{TSMB}.

 In the standard HMC algorithm we used even-odd preconditioning, which
 in presence of a twisted mass is only a slight modification of the
 standard preconditioning technique \cite{DEGRAND}.
 We give the relevant equations in appendix~\ref{app.hmc} of this paper.
 As a subsequent improvement of the algorithm, we implemented the idea
 of ref.~\cite{HASENBUSCH} and used shifted fermion matrices to
 ``precondition'' the original fermion matrix.
 These shifted matrices are treated by introducing additional
 pseudo-fermion fields.
 In the shifted fermion matrix we simply used larger values of the
 twisted mass parameter than the value of $\mu$ that is to be simulated.
 Using two pseudo-fermion fields we experienced a substantial
 improvement of the HMC algorithm by at least a factor of two.
 The addition of a third pseudo-fermion field gave only another 10-20\%
 improvement.
 Again we list the relevant equations, how the shifted matrices are
 implemented, in appendix \ref{app.hmc}.
 As a further algorithmic trick we used the Sexton-Weingarten leap-frog
 integrator as proposed in ref.~\cite{SEXTON}.

 Our alternative algorithm, the TSMB algorithm \cite{TSMB}, is based
 on the multi-boson representation of the fermion determinant
 \cite{LUSCHER}.
 Optimized polynomial approximations are used, both in the first
 update step and in the second global accept-reject correction step, for
 reproducing the dynamical effect of fermions on the gauge field.
 We apply high order least-square optimization and obtain the necessary
 polynomials using high precision arithmetics \cite{GEBERT-MONTVAY}.
 Concerning the optimization of TSMB for QCD see, for instance,
 \cite{QQ+Q1}.

 A useful improvement of the TSMB update algorithms can again be
 achieved by even-odd preconditioning.
 This can be implemented in TSMB for twisted mass quarks along the
 lines of ref.~\cite{JEGERLEHNER}.
 For the even-odd preconditioning of the TSMB update the flavour
 indices of the quark fields have to be kept.
 This means that the multi-boson fields have 24 components per lattice
 site (2 for flavour, 3 for colour and 4 for Dirac spinor indices).
 Correspondingly, the polynomials are approximating the function
 $x^{-\half}$ as in the case of a single Dirac flavour with untwisted
 quark mass.
 We give some more details of our even-odd implementation of the TSMB
 algorithm in appendix \ref{app.tsmb}.

 In the region of light quarks an important part of the numerical effort
 has to be spent on equilibrating the gauge configuration in a new
 simulation point.
 This is particularly relevant in studies of the phase structure where
 many different points in the parameter space have to be investigated.
 In case of TSMB the equilibration time is substantially longer than
 the autocorrelation of relevant physical quantities in equilibrium:
 on our lattices equilibration can take ten or more times the
 autocorrelation time of the plaquette observable.
 The autocorrelation times in equilibrium themselves are similar but
 most of the time by factors of 2-3 shorter with our twisted quark
 masses than with untwisted quark masses of similar magnitude.
 For an approximate formula of the computational cost see
 ref.~\cite{QQ+Q2}.
 
 The use of two different optimized update algorithms was very helpful
 in checking our results.
 We did not try to obtain a precise performance comparison.
 Qualitatively, we did not see a noticeable difference in the speed
 once equilibrium was reached, but the HMC algorithm with multiple pseudofermion
 fields (MPHMC) turned out to be faster in the equilibration process.
 In particular, crossing the transition region below and above the
 metastability region is faster with MPHMC.
 Nevertheless, the extension in $\kappa$ of the metastability region is the same
 with both algorithms.

 The data used for preparing the figures in this publication were
 obtained with MPHMC, except for the upper four panels in
 fig.~\ref{metastable}, which were obtained with TSMB.
 The thermal cycles in fig.~\ref{cycles} were only run with MPHMC.
 In the other figures the results of the TSMB runs, whenever
 performed, were always consistent within errors with the shown MPHMC
 results.

%%%%%%%%%%%%%%%%%%%%%%%%%%%%%%%%%%%%%%%%%%%%%%%%%%%%%%%%%%%%%%%%%%%%%%%%
\section{Numerical results}\label{sec3}

 In this section we give our numerical evidence for the phenomenon of
 metastability mentioned in the introduction.
 As a first step and for an orientation we have investigated thermal
 cycles in the hopping parameter $\kappa$.
 We then discuss metastable states in the plaquette expectation value.
 Finally we determine quantities such as the pion mass and the
 untwisted PCAC quark mass in the metastable branches in order to obtain
 a picture of the physical properties in the different states.
 In most cases we perform the simulations at a twisted mass $a\mu=0.01$
 but in a few cases we also put $a\mu=0$, which is possible on the
 lattice sizes we consider.

%%%%%%%%%%%%%%%%%%%%%%%%%%%%%%%%%%%%%%%%%%%%%%%%%%%%%%%%%%%%%%%%%%%%%%%%
\subsection{Thermal cycles}\label{sec3.1}

 We started our investigation of the phase diagram of zero temperature
 lattice QCD by performing thermal cycles in $\kappa$ while keeping
 fixed $\beta=5.2$ and the value of the twisted mass parameter $a\mu$.
 These cycles are performed such that a starting value of
 $\kappa_{\mathrm{start}}$ is chosen and then $\kappa$ is incremented,
 without performing further intermediate thermalization sweeps, until a
 final value of $\kappa_{\mathrm{final}}$ is reached.
 At this point the procedure is reversed and $\kappa$ is decremented
 until the starting value $\kappa_{\mathrm{start}}$ is obtained back.
 At each value of $\kappa$ $150$ configurations are produced and
 averaged over.

%%%%%%%%%%%%%%%%%%%%%%%%%%%%%%%%%%%%%%%%%%%%%%%%%%%%%%%%%%%%%%%%%%%%%%%%
\begin{figure}
\vspace*{-8mm}
\begin{center}
\epsfig{file=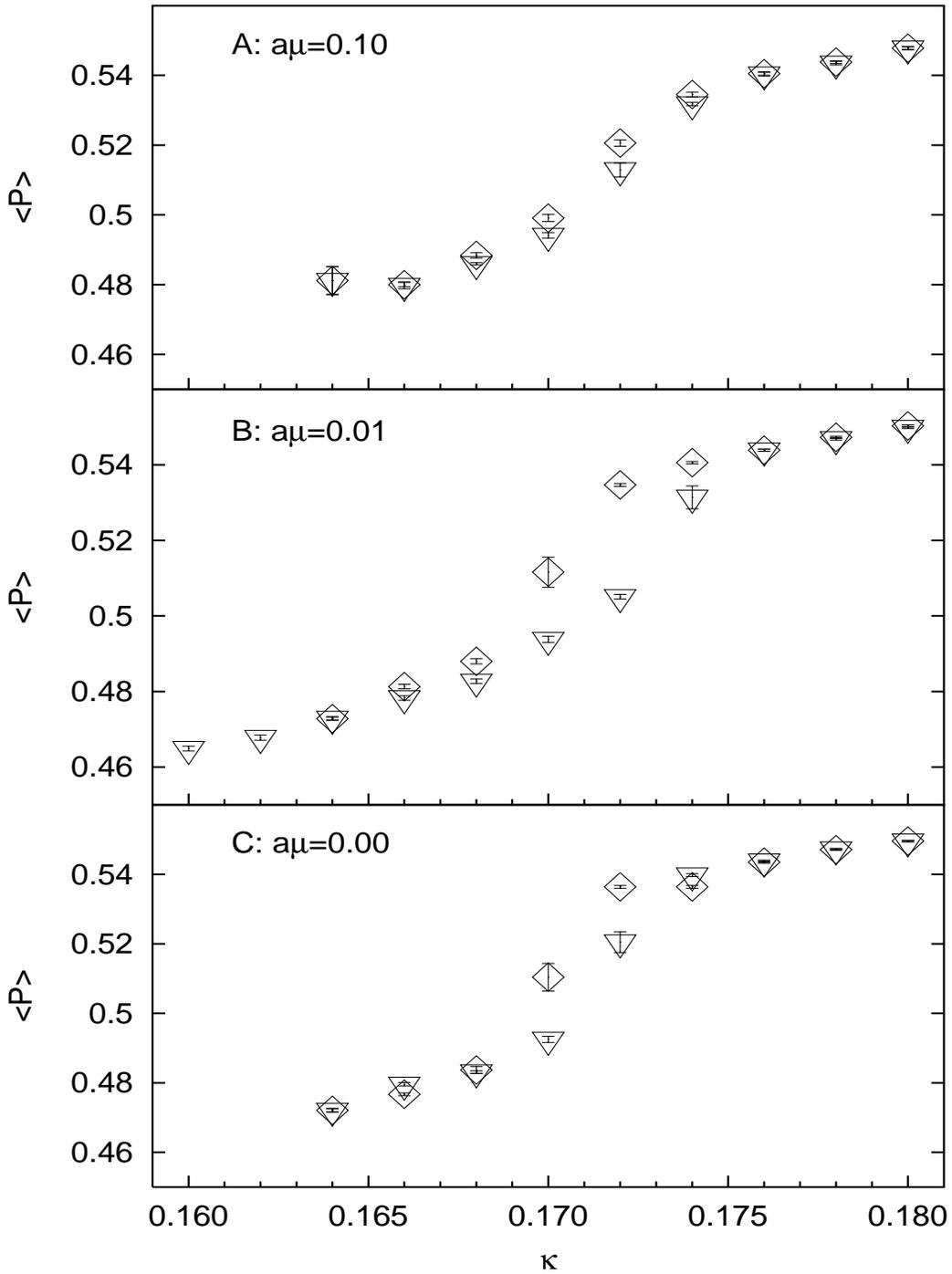,
        width=147mm,height=190mm}
\end{center}
\vspace*{-8mm}
\begin{center}
\parbox{12cm}{\caption{
 Thermal cycles in $\kappa$ on $8^3\times 16$ lattices at $\beta=5.2$.
 The plaquette expectation value is shown for:
 $a\mu=0.1$ (A); $a\mu=0.01$ (B); $a\mu=0$ (C).
 The triangles ($\triangledown$) refer to increasing $\kappa$-values, the diamonds 
 ($\Diamond$) to decreasing ones.
\label{cycles}}}
\end{center}
\end{figure}
%%%%%%%%%%%%%%%%%%%%%%%%%%%%%%%%%%%%%%%%%%%%%%%%%%%%%%%%%%%%%%%%%%%%%%%%

 In fig.~\ref{cycles} we show three such thermal cycles, performed at
 $a\mu=0$, $a\mu=0.01$ and $a\mu=0.1$ from bottom to top.
 In the cycles signs of hysteresis effects can be seen for $a\mu=0$ and
 $a\mu=0.01$ while for the largest value of $a\mu=0.1$ such effects are
 hardly visible.
 Hysteresis effects in thermal cycles {\em may be} signs of the
 existence of a first order phase transition.
 However, they should only be taken as first indications.
 Nevertheless, they provide most useful hints for further studies to
 search for metastable states.

%%%%%%%%%%%%%%%%%%%%%%%%%%%%%%%%%%%%%%%%%%%%%%%%%%%%%%%%%%%%%%%%%%%%%%%%
\subsection{Metastability}\label{sec3.2}

\begin{figure}
\begin{center}
\begin{minipage}[c]{1.0\linewidth}  
                    % CHANGE THIS NUMBER FOR ADJUSTING TOTAL SIZE
 \begin{minipage}[c]{0.50\hsize}
  \includegraphics[angle=-90,width=.95\hsize]{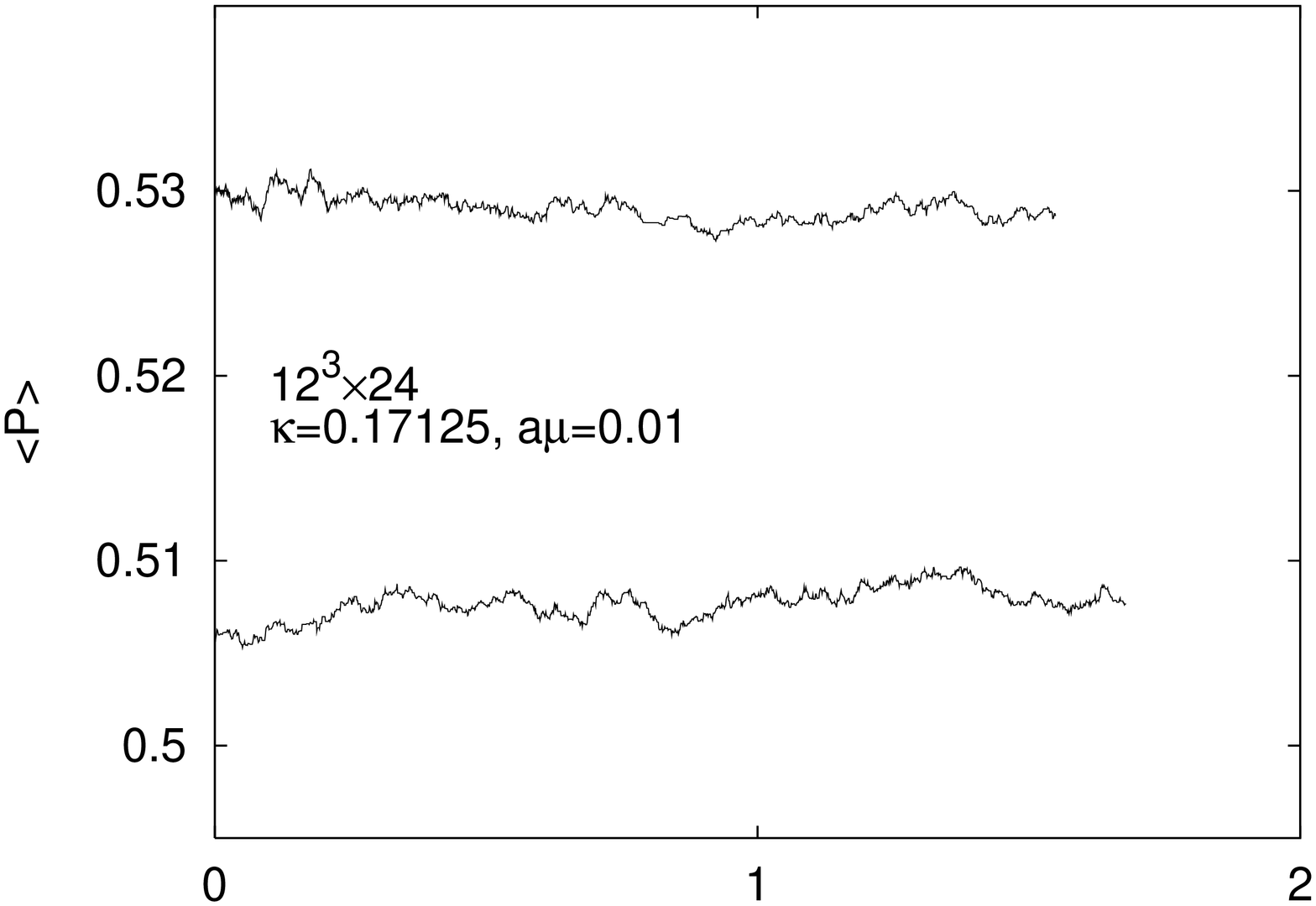}
  \includegraphics[angle=-90,width=.95\hsize]{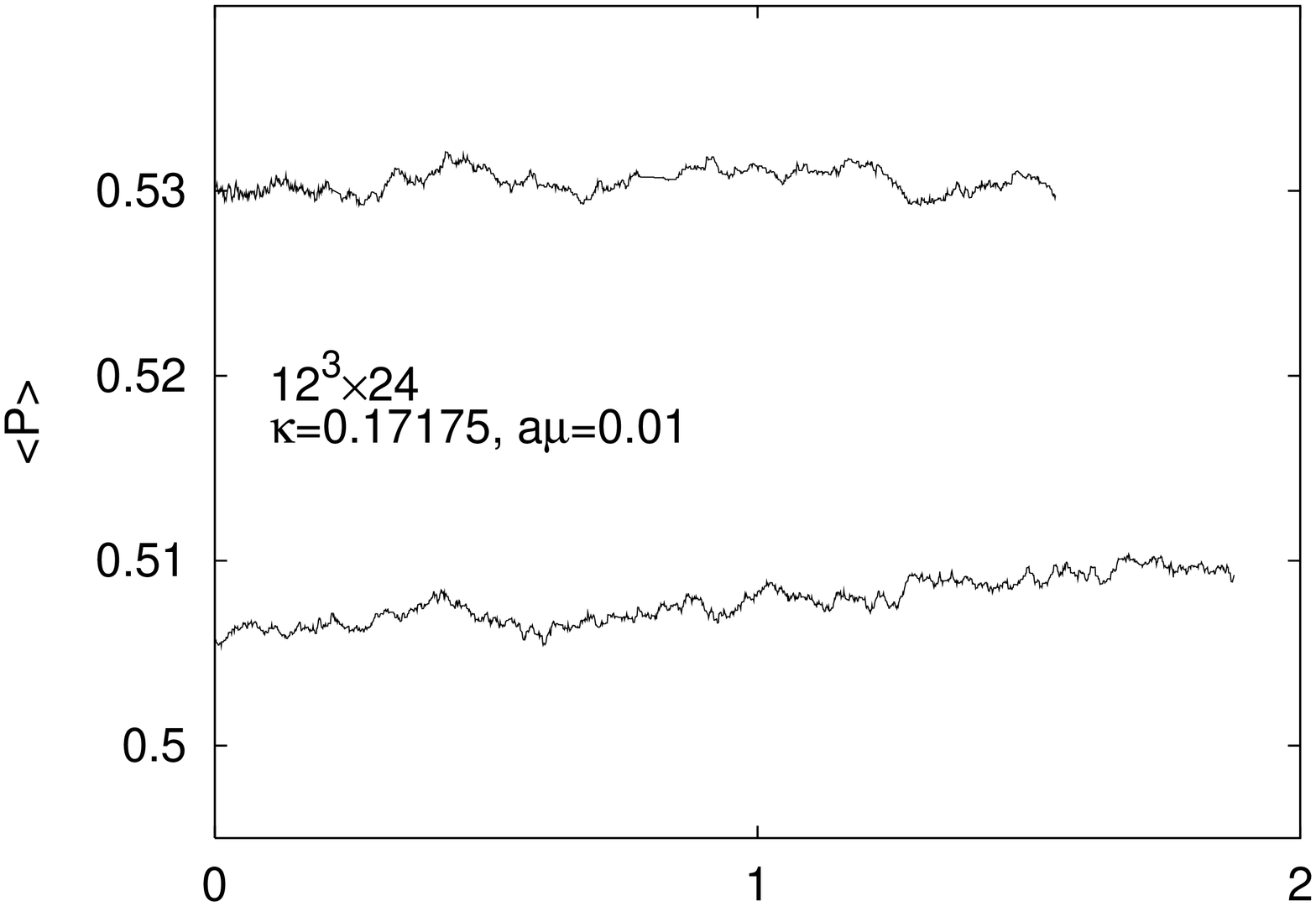}
  \includegraphics[angle=-90,width=.95\hsize]{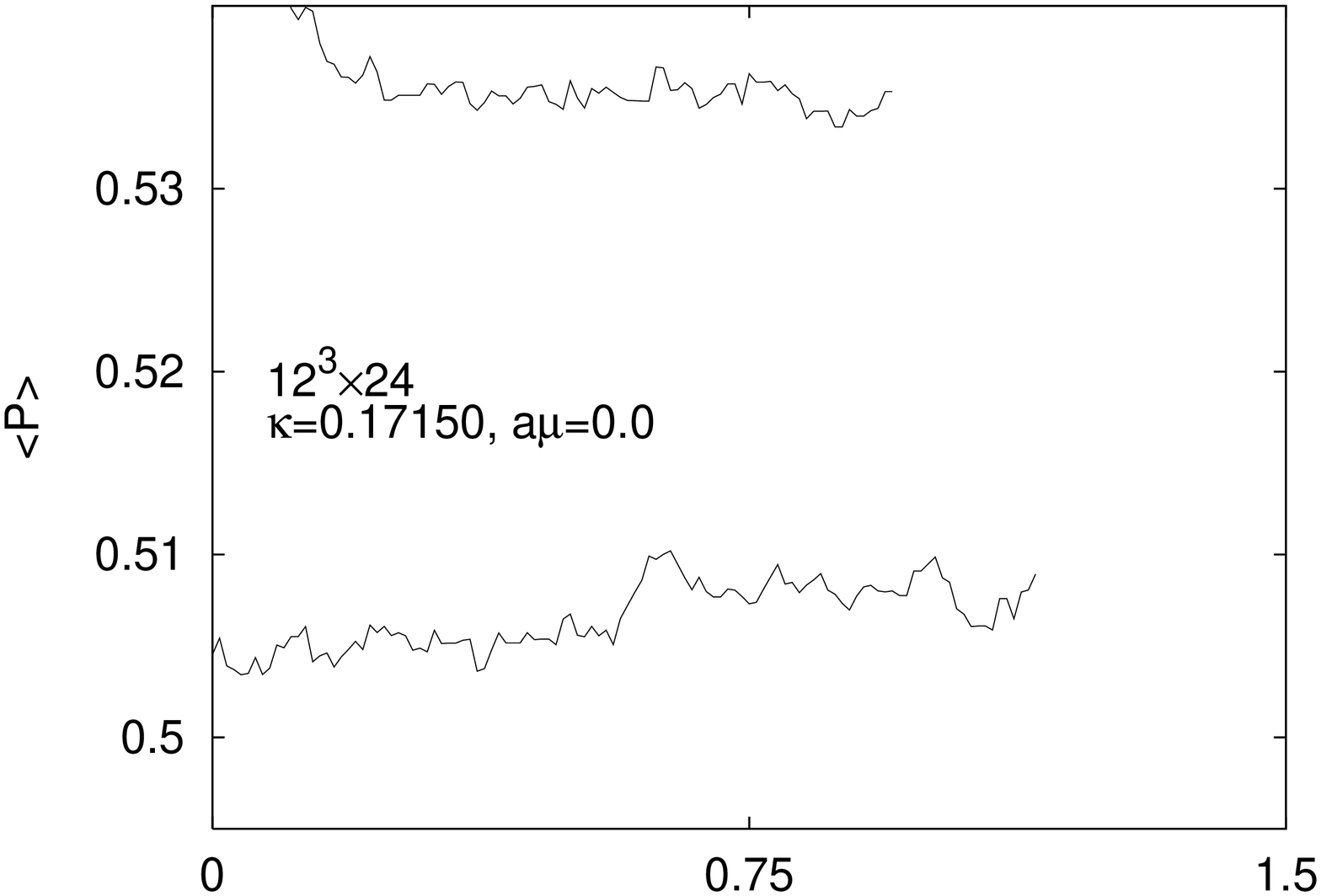}
 \end{minipage}\hfill
 \begin{minipage}[c]{0.50\hsize}
  \includegraphics[angle=-90,width=.95\hsize]{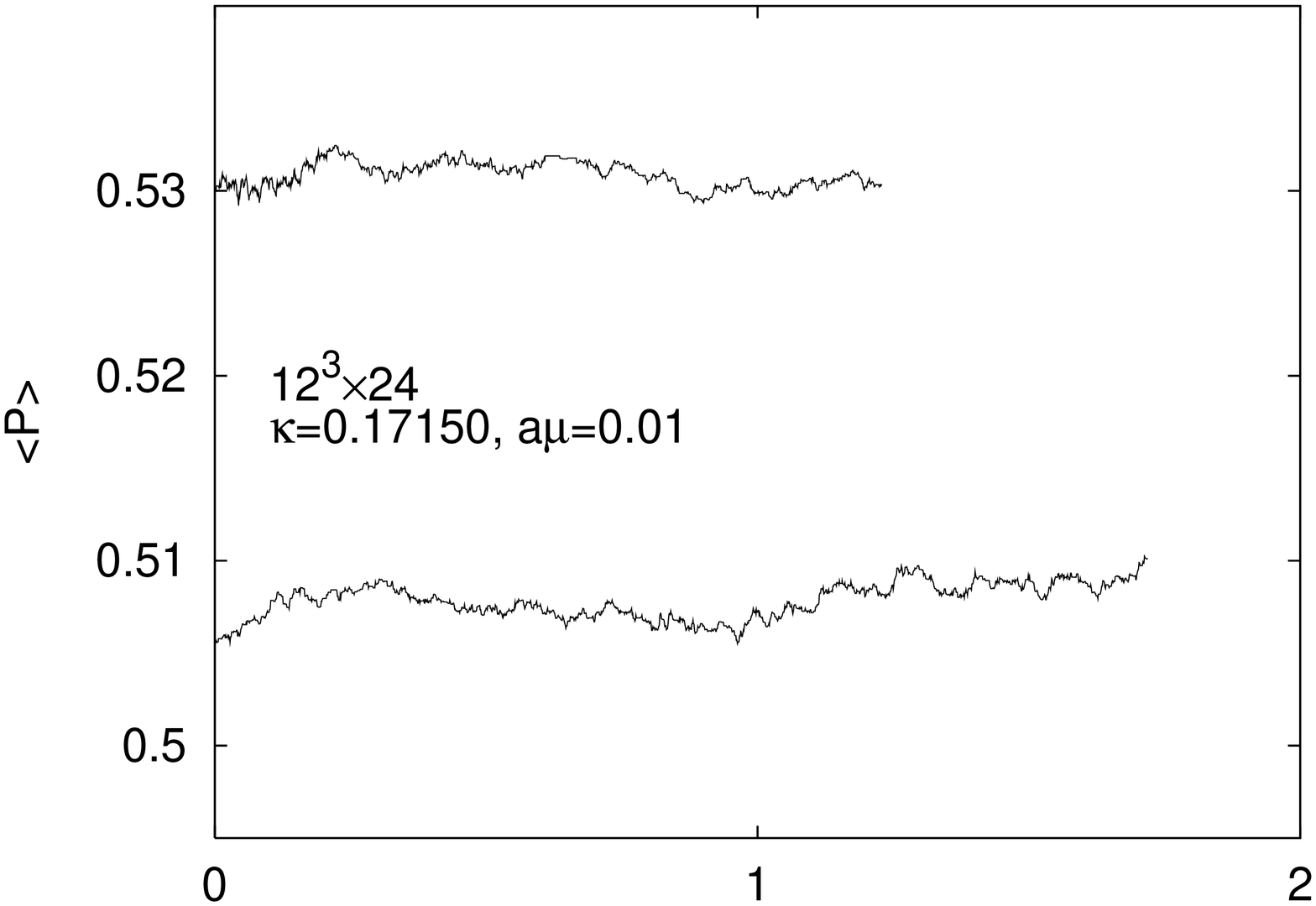}
  \includegraphics[angle=-90,width=.95\hsize]{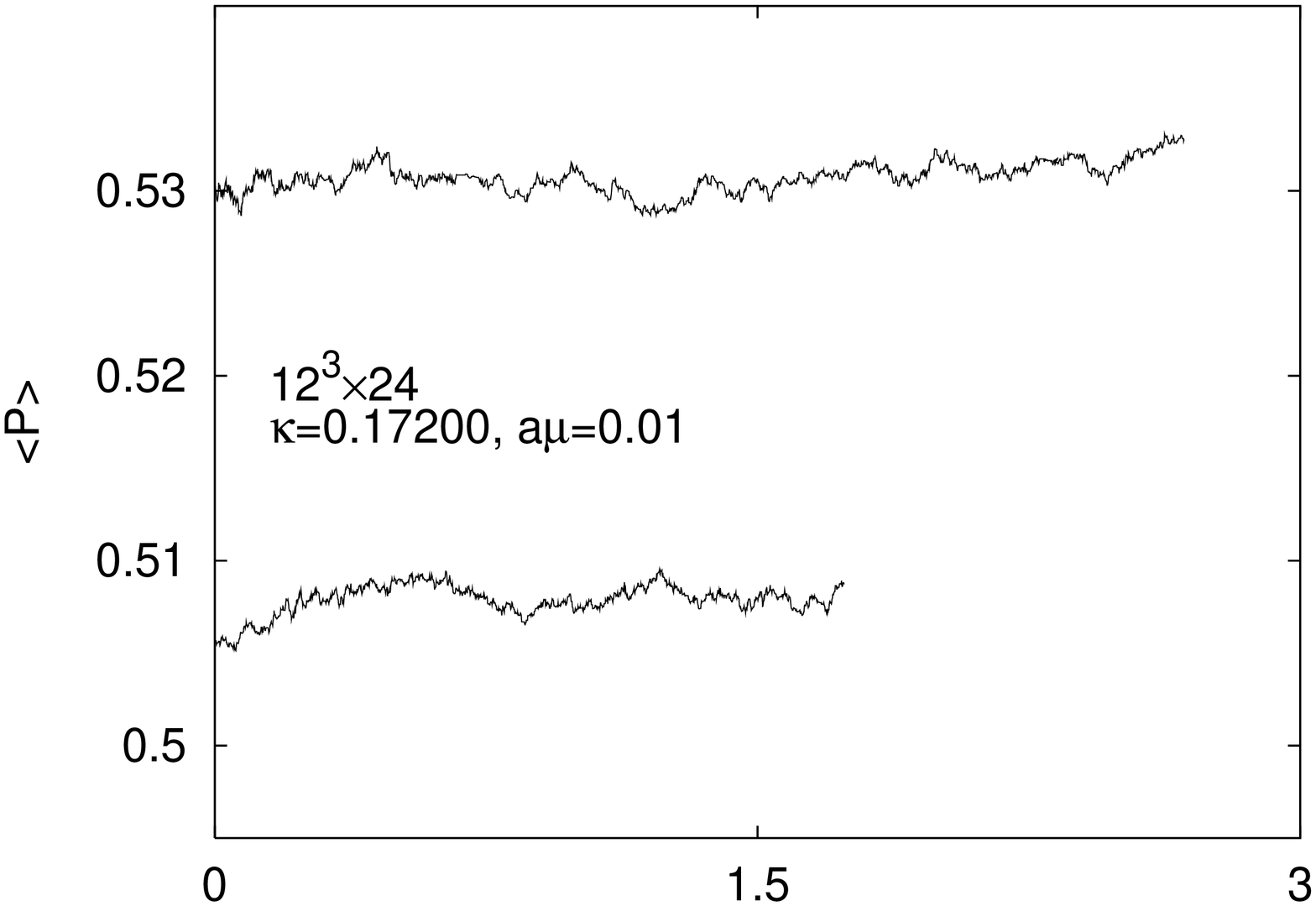}
  \includegraphics[angle=-90,width=.95\hsize]{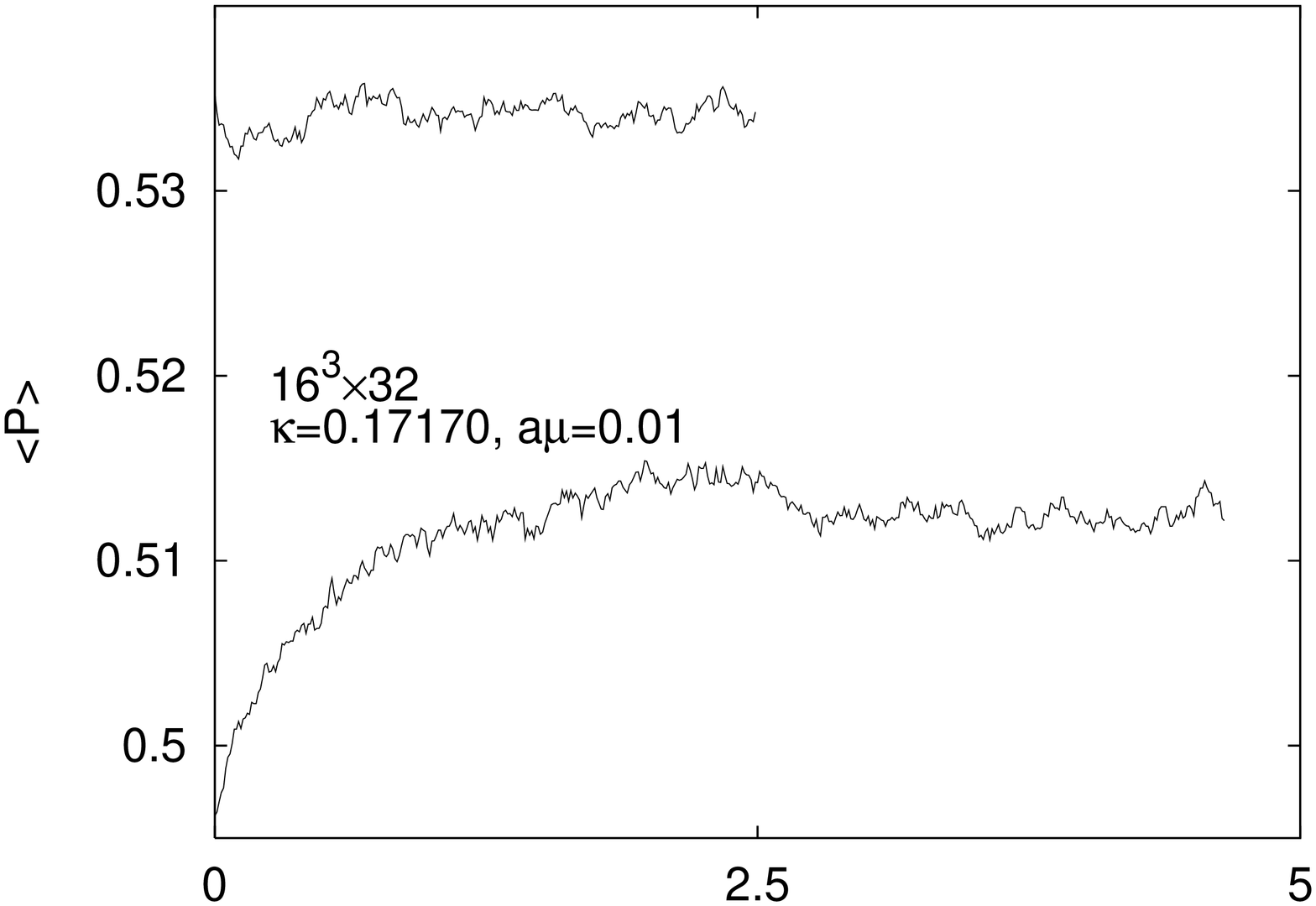}
 \end{minipage}
\end{minipage}
\end{center}
\begin{center}
\parbox{16cm}{\caption{
 Metastable states at $\beta=5.2$.
 The number of sweeps is given in thousands.
 The lattice size is $12^3 \times 24$, except for the right panel
 in the bottom line where it is $16^3 \times 32$.
 The twisted mass is $a\mu=0.01$, except for the left panel in the bottom
 line where it is $a\mu=0$.
\label{metastable}}}
\end{center}
\end{figure}

 Guided by the results from the thermal cycles, we next performed
 simulations at fixed values of $a\mu$ and $\kappa$, starting with
 ordered and disordered configurations, staying again at $\beta=5.2$.
 In fig.~\ref{metastable} we show the Monte Carlo time evolution of the
 plaquette expectation value, in most cases on a $12^3 \times 24$
 lattice.
 For several values of $\kappa$ we find coexisting branches with
 different average values of the plaquette.
 The gap (the ``latent heat'') appears to be rather large.
 At $\kappa=0.1717$ we show the history of the plaquette expectation
 value also on a larger ($16^3 \times 32$) lattice.
 It seems that the gap in the plaquette expectation value does not
 depend much on the lattice size, suggesting that the metastability we
 observe here is not a finite volume effect.
 In most cases the twisted mass is $a\mu=0.01$, except for the picture
 left in the bottom line where it is $a\mu=0$.

%%%%%%%%%%%%%%%%%%%%%%%%%%%%%%%%%%%%%%%%%%%%%%%%%%%%%%%%%%%%%%%%%%%%%%%%
%%%%%%%%%%%%%%%%%%%%%%%%%%%%%%%%%%%%%%%%%%%%%%%%%%%%%%%%%%%%%%%%%%%%%%%%

 The lifetime of a metastable state, i.e.~the time before a tunneling
 to the stable branch occurs, depends on the algorithm used.
 In fact, one may wonder, whether the appearance of the metastable
 states seen in fig.~\ref{metastable} may not be purely an artefact of
 our algorithms.
 We cannot completely exclude this possibility but we believe it is very
 unlikely: we employed two very different kinds of algorithms in our
 simulations as explained in section~\ref{sec2.2}.
 We observe the metastable states with both of them.
 We also inter-changed configurations between the two algorithms:
 a configuration generated with the algorithm A was iterated further
 with algorithm B and vice versa.
 We find that in such situations the plaquette expectation value remains
 in the state where it has been before the interchange of configurations
 took place.
 In addition, as we shall see below, the two states can be characterized
 by well defined and markedly different values of basic quantities.
 We therefore conclude that the metastable states are a generic
 phenomenon of lattice QCD in our formulation.

%%%%%%%%%%%%%%%%%%%%%%%%%%%%%%%%%%%%%%%%%%%%%%%%%%%%%%%%%%%%%%%%%%%%%%%%
\subsection{Pion and quark masses}\label{sec3.3}

 By selecting separately configurations with high and with low
 plaquette expectation value, we measured the pion mass and the
 untwisted PCAC quark mass to study the physical properties in the two
 metastable states.

 We obtained the pseudo-scalar (``pion'') mass from suitable correlation
 functions.
 These are constructed from the standard composite fields defined in
 terms of the fields $\bar\psi$ and $\psi$ in eq.~(\ref{phaction}):
\begin{eqnarray}
S^0(x) & = & \bar{\psi}(x)\psi(x), \hspace*{6em}
P^\alpha(x)  = \bar{\psi}(x)\gamma_5\frac{\tau_\alpha}{2}\psi(x) ,  
\nonumber \\
A_\mu^\alpha(x) & = &
           \bar{\psi}(x)\gamma_\mu\gamma_5\frac{\tau_\alpha}{2}\psi(x) ,
\hspace*{3em}
V_\mu^\alpha(x) = \bar{\psi}(x)\gamma_\mu\frac{\tau_\alpha}{2}\psi(x)\ .
\label{operators}
\end{eqnarray}
 Here $\tau_\alpha,\; \alpha=1,2,3$ are the usual Pauli-matrices
 in isospin space.
 The corresponding composite fields in terms of the quark fields $\chi$
 and $\bar{\chi}$ of eq.~(\ref{tmaction}) are then given by the
 transformation in eq.~({\ref{transformation}).
 For instance, for $\alpha=1,2$ (``charged pions'') the pseudo-scalar
 density has the same form in the $\chi$-basis as in the $\psi$-basis.
 Therefore, the mass of the charged pions can be extracted from
 correlators in the $\chi$-basis in the usual way.
 The charged axial vector and vector currents are rotated into each other
 by the angle $\omega$ in such a way that at $\omega=\pi/2$ they
 are interchanged.
 (For more details see the literature,
 e.g.~\cite{TMQCD,FREZZOTTI-ROSSI}.)

 Besides the pion mass, we measured the PCAC quark mass from the
 axial vector current in the $\chi$-basis:
\begin{equation}\label{mpcac}
m_\chi^{\mathrm{PCAC}} \equiv
\frac{\langle\partial_\mu^*\bar\chi\gamma_\mu\gamma_5
\frac{\tau^\pm}{2}\chi(x)\;  \hat O^\mp(y)\rangle}
{2\langle\bar\chi\frac{\tau^\pm}{2}\gamma_5\chi(x)\; 
\hat O^\mp(y)\rangle} \ .
\end{equation}
 Here $\hat O^\mp$ is a suitable operator that we have chosen to be
 the pseudo-scalar density
 $\hat O^\mp = \bar\chi\frac{\tau^\mp}{2}\gamma_5\chi(x)$,
 $\partial_\mu^*$ is the lattice backward derivative defined as usual
 and $\tau^\pm = \tau_1\pm i\tau_2$.
 One can show that in the limit $a\to 0$ the quantity
 $m_\chi^{\mathrm{PCAC}}$ is asymptotically proportional, through
 finite renormalization constants, to $\tilde{m}$.

 In fig.~\ref{pionmass} we show the pion mass squared in lattice units as
 function of $(2\kappa)^{-1}$.
 We observe that the pion mass is rather large and the most striking
 effect in the graph is that it can have two different values at the
 same $\kappa$.
 If we consider the quark mass $m_\chi^{\mathrm{PCAC}}$ in
 fig.~\ref{quarkmass}, we see that in the states with a low plaquette
 expectation value the mass is positive while for high values of the
 plaquette expectation it is negative.
 These quark masses with opposite sign coexist for some values
 of~$\kappa$.
 Plotting the pion mass versus $m_\chi^{\mathrm{PCAC}}$ one obtains
 fig.~\ref{pionmass_vs_quarkmass}.

%%%%%%%%%%%%%%%%%%%%%%%%%%%%%%%%%%%%%%%%%%%%%%%%%%%%%%%%%%%%%%%%%%%%%%%%
\begin{figure}
\vspace*{-60mm}
\begin{center}
\epsfig{file=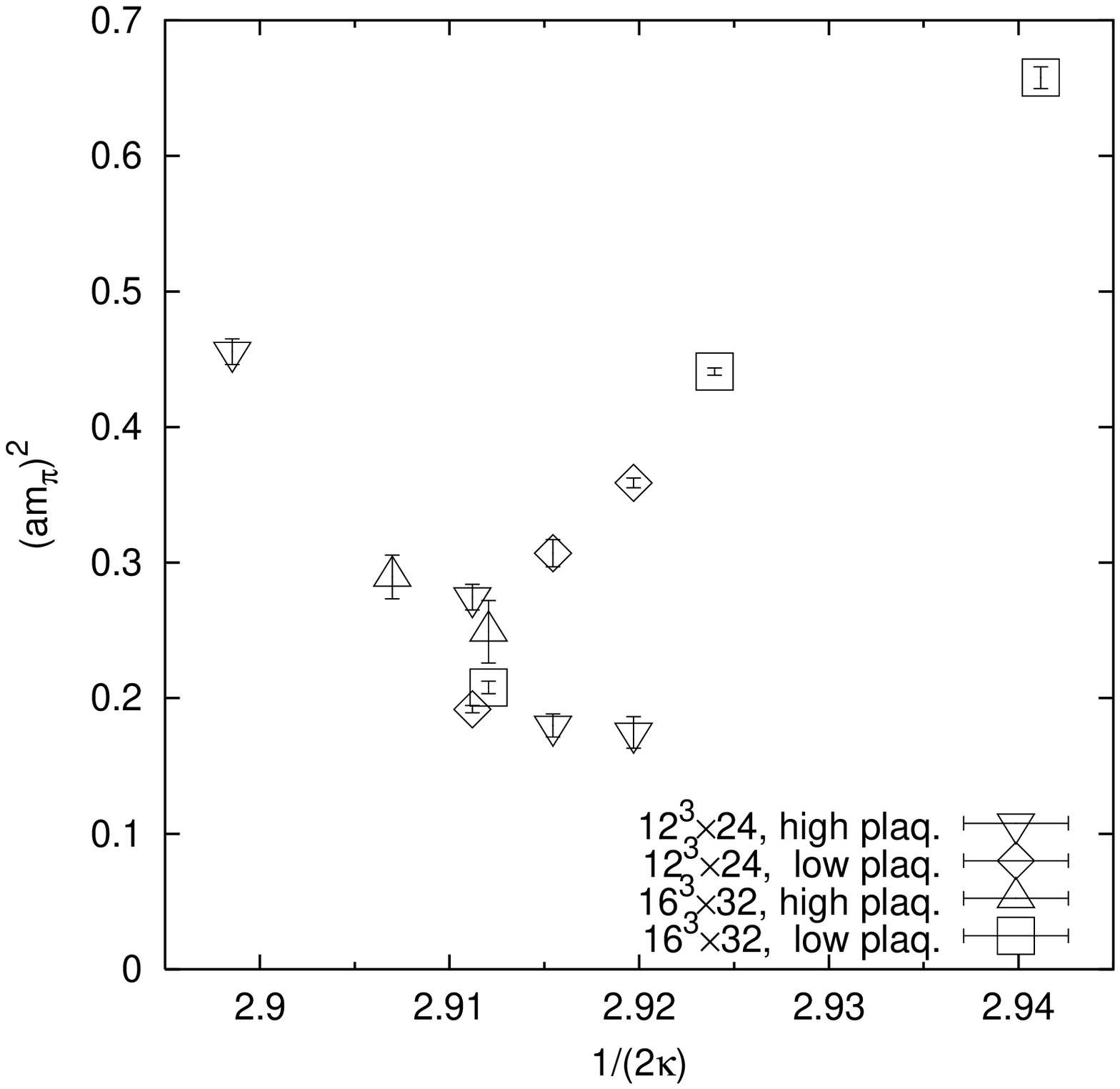,
        width=140mm,height=210mm}
\end{center}
\begin{center}
\parbox{12cm}{\caption{
 The pion mass squared in lattice units on two lattice sizes measured
 separately on configurations in the two metastable states.
 These runs were made at $\beta=5.2$ and $a\mu=0.01$.
\label{pionmass}}}
\end{center}
\end{figure}
%%%%%%%%%%%%%%%%%%%%%%%%%%%%%%%%%%%%%%%%%%%%%%%%%%%%%%%%%%%%%%%%%%%%%%%%

%%%%%%%%%%%%%%%%%%%%%%%%%%%%%%%%%%%%%%%%%%%%%%%%%%%%%%%%%%%%%%%%%%%%%%%%
\begin{figure}
\vspace*{-40mm}
\begin{center}
\epsfig{file=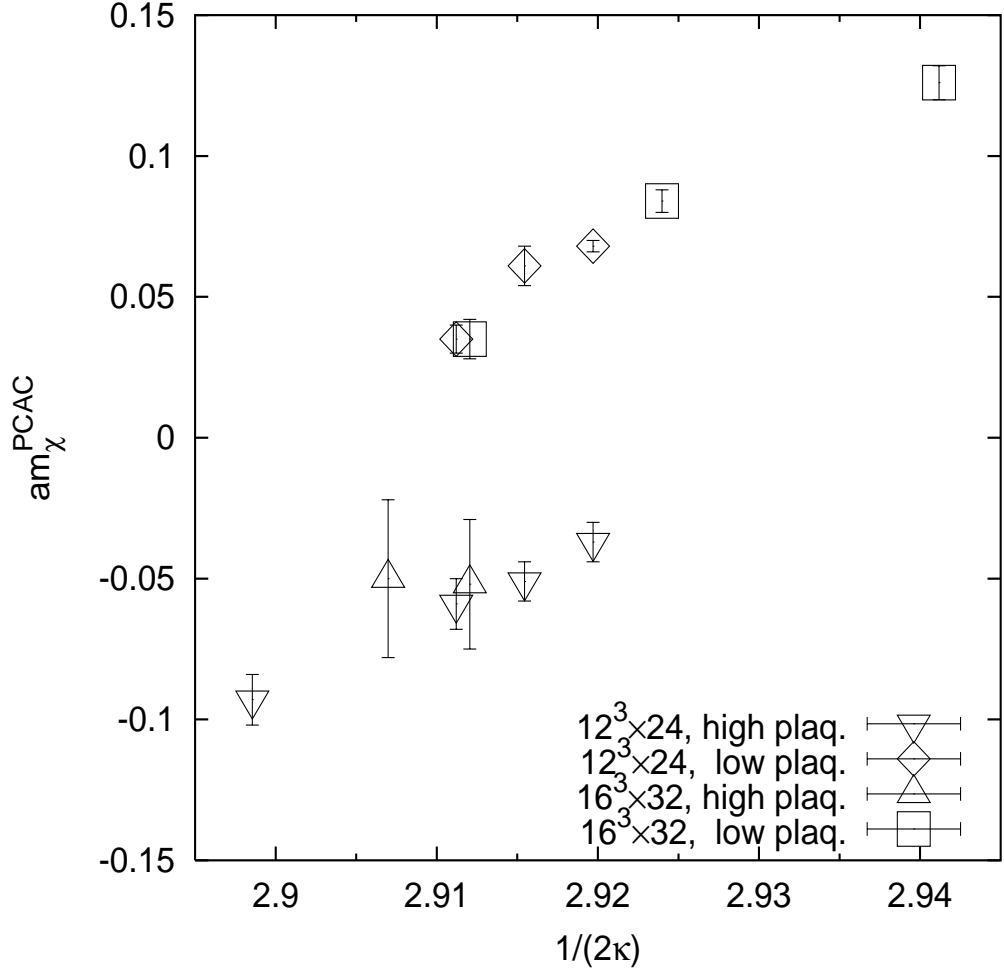,
        width=140mm,height=210mm}
\end{center}
\begin{center}
\parbox{12cm}{\caption{
 The quark mass in lattice units $m_\chi^{\mathrm{PCAC}}$ as defined in
 eq.~(\protect\ref{mpcac}) on two lattice sizes measured separately on
 configurations in the two metastable states.
 The values of $\beta=5.2$ and $a\mu=0.01$ are fixed.
\label{quarkmass}}}
\end{center}
\end{figure}
%%%%%%%%%%%%%%%%%%%%%%%%%%%%%%%%%%%%%%%%%%%%%%%%%%%%%%%%%%%%%%%%%%%%%%%%

%%%%%%%%%%%%%%%%%%%%%%%%%%%%%%%%%%%%%%%%%%%%%%%%%%%%%%%%%%%%%%%%%%%%%%%%
\begin{figure}
\vspace*{10mm}
\begin{center}
\epsfig{file=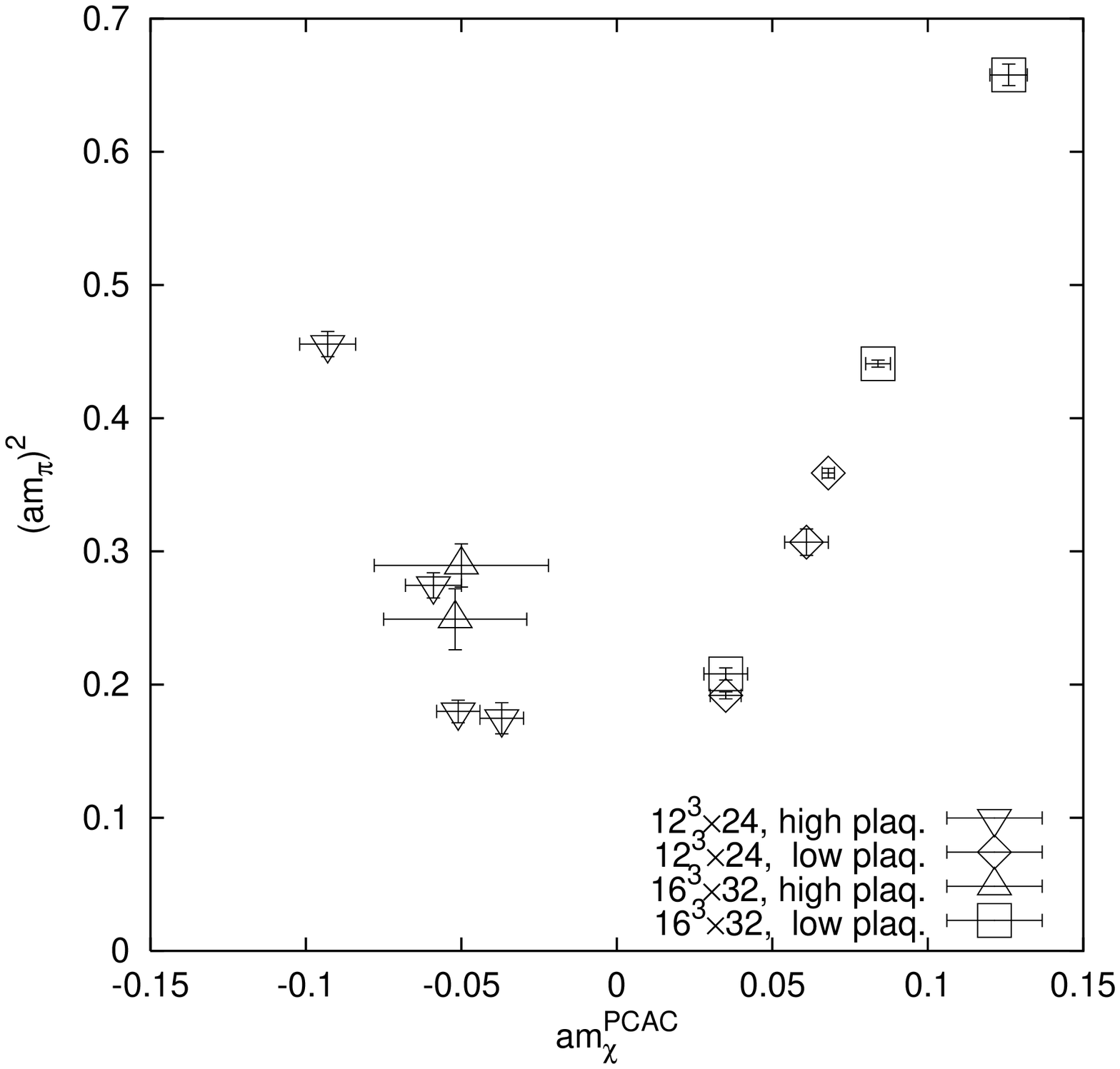,
        width=150mm,height=200mm,angle=-90}
\end{center}
\begin{center}
\parbox{15cm}{\caption{
 The pion mass squared in lattice units from fig.~\protect\ref{pionmass}
 plotted against the untwisted PCAC quark mass in
 fig.~\protect\ref{quarkmass}.
\label{pionmass_vs_quarkmass}}}
\end{center}
\end{figure}
%%%%%%%%%%%%%%%%%%%%%%%%%%%%%%%%%%%%%%%%%%%%%%%%%%%%%%%%%%%%%%%%%%%%%%%%

 Figs.~\ref{metastable}-\ref{quarkmass} clearly reveal that for small
 enough $\mu$ metastabilities show up in the quantities we have
 investigated, such as $m_\pi$, $m_\chi^{\mathrm{PCAC}}$ and the average
 plaquette, if $m_0$ is close to its critical value.
 What ``small enough $\mu$'' means is likely to change with $\beta$.
 Simulations at larger values of $\beta$ are in progress.
 As a matter of fact, when $m_0$ is significantly larger (smaller) than
 $m_{\rm crit}$ we find $m_\chi^{\mathrm{PCAC}}$ to be positive (negative) and no
 signal of metastabilities.
 The remark that metastabilities take place for $m_0$ close to its
 critical value will be important both in sect.~\ref{sec4.1} to
 understand why they affect also a purely gluonic observable such as the
 plaquette and in sect.~\ref{sec4.3}, where it leads to a plausible
 explanation of the observed metastability phenomena in terms of
 spectral properties of the lattice tmQCD Dirac matrix (suppression of
 the ``eigenvalue cloud crossing'' phenomenon by the fermionic
 determinant).

 The remarks in sect.~\ref{sec4.1} may provide further insight also on
 the similar metastability phenomena reported in \cite{JLQCD}
 for the $N_f=3$ untwisted Wilson theory and on the reason why they 
 ``disappear'' when changing the gluonic action or details,
 e.g.~$c_{SW}$-value, of the fermionic action.
 A possible reason is that these changes might shift the range of
 $m_0$ where metastabilities appear to values where no data are yet
 available.

%%%%%%%%%%%%%%%%%%%%%%%%%%%%%%%%%%%%%%%%%%%%%%%%%%%%%%%%%%%%%%%%%%%%%%%%
\section{Physical interpretation}\label{sec4}

 The observed strong metastabilities discussed in the previous section
 clearly suggest that we are working either directly at a first order
 phase transition or at least very close to it such that we see the
 remnants of a close-by first order phase transition.
 With the present data we cannot really differentiate between these
 two scenarios and in the following we will therefore discuss both of
 them. 

%%%%%%%%%%%%%%%%%%%%%%%%%%%%%%%%%%%%%%%%%%%%%%%%%%%%%%%%%%%%%%%%%%%%%%%%
\subsection{Jump in the plaquette and chiral symmetry breaking}
\label{sec4.1}

 Generally speaking, a jump in the plaquette as seen in our data
 can arise owing to the lack of chiral symmetry for chirally
 non-invariant formulations of lattice QCD.
 The argument relies on the key observation that, when working with
 chirally twisted Wilson fermions, there are two distinct sources of
 chirality breaking.
 The first source is the combination of the untwisted Wilson and mass
 terms
\begin{eqnarray}
\label{massterms}
\bar{\chi}\, M[U]\, \chi 
& = & a^4\sum_x \left\{\bar\chi(x)\Big(\frac{4r}{a}
          + m_0\Big) \chi(x) \right. 
\nonumber \\
& - & \left. \frac{r}{2a}\bar\chi(x) \sum_{\mu=1}^4 \Big(
        U(x,\mu) \chi(x+a\hat\mu) -
U^\dagger(x-a\hat\mu,\mu)\chi(x-a\hat\mu)\Big)\right\}\; .
\end{eqnarray}
 The second source of chirality breaking is the twisted mass term
 $\mu\bar{\chi} i\gamma_5\tau_3 \chi$.
 As pointed out in section~\ref{sec2}, one may trade the bare
 parameters $m_0$ and $\mu$ in eq.~(\ref{tmaction}) for the equivalent
 bare parameters $m_q$ and $\omega$ of eq.~(\ref{phaction}).
 The latter are best suited to discuss the connection with continuum
 QCD physics, as $\omega$ is an unphysical parameter, while $m_q$
 represents the bare quark mass.
 Assuming spontaneous chiral symmetry breaking in infinite volume,
 the pion mass squared is expected to vanish linearly in $m_q$ (up
 to lattice artifacts) as $m_q \to 0$.
 Moreover in the continuum limit the physical scalar condensate is
 expected to show a discontinuity and changes sign as $m_q$ passes
 through zero:
\begin{equation}
\lim_{m_q\rightarrow 0^{+}} \langle [\bar{\psi}\psi]_{\rm R}\rangle =
-\lim_{m_q\rightarrow 0^{-}} \langle [\bar{\psi}\psi]_{\rm R} \rangle\ne
0\; ,
\label{jumppbarp}
\end{equation}
 where by $[\bar{\psi}\psi]_{\rm R}$ we mean the appropriately
 subtracted and renormalized scalar density.
 We recall that for $\omega \neq 0$ this is a non-trivial linear
 combination of $\bar{\chi}\chi$, $\bar{\chi} i\gamma_5\tau_3 \chi$ and
 the constant field (see below for details). 

 In order to make contact with the observed metastability phenomena
 in the regime of spontaneous chiral symmetry breaking, two further
 remarks are important:
\begin{enumerate}

\item at non-zero lattice spacing the twisted mass term
 $\mu \bar{\chi} i\gamma_5\tau_3 \chi$ induces the twisted condensate
 $\langle [\bar{\chi} i\gamma_5\tau_3 \chi]_{\rm R} \rangle$, while
 the untwisted mass terms $\bar{\chi}M[U]\chi$ of eq.~(\ref{massterms}) determines
 the untwisted condensate $\langle [\bar{\chi}\chi]_{\rm R}\rangle$.

\item the local plaquette field
\begin{equation} \label{LOCPLAQ}
\phi(x) \equiv \frac{1}{12} \sum_{\mu \neq \nu} \frac{1}{3}
{\rm tr} [ U_\mu(x) U_\nu(x+a\hat\mu) U^\dagger_\mu(x+a\hat\nu)
U^\dagger_\nu(x) ]
\end{equation}
 admits on the basis of lattice symmetries an operator expansion
 of the form
\begin{equation}
\label{PHIREPR} 
\phi(x) = \Big{[} b_0 {\mathds 1}+ b_{4g}\, a^4 F \cdot F \Big{]} + 
b_3\, a^3 [\bar{\chi} \chi]_{\rm sub}
+ b_{4}\, a^4 \mu [\bar{\chi} i\gamma_5 \tau_3 \chi]_{\rm sub}
\; + \;  {\rm O}(a^5) \, ,
\end{equation}
 with $[...]_{\rm sub}$ denoting a subtracted, multiplicatively 
 renormalizable, operator and $F$ the continuum gauge field strength
 tensor. The plaquette expectation value $P(r,am_0,a\mu)$ can be
 correspondingly written in the form 
\begin{eqnarray}
\label{PREPR}
P(r,am_0,a\mu) & = & \Big{[} b_0 +
b_{4g}\, a^4 \langle F \cdot F \rangle_{(r,am_0,a\mu)}
\Big{]} + b_3\, a^3 \langle [\bar{\chi} \chi]_{\rm sub}
\rangle_{(r,am_0,a\mu)} \nonumber \\[0.5em]
&  & + \; b_{4}\, a^4 \mu \langle [\bar{\chi} i\gamma_5 \tau_3
\chi]_{\rm sub} \rangle_{(r,am_0,a\mu)}  \; + \;  {\rm O}(a^5) \, .
\end{eqnarray}
 The important point about the representation~(\ref{PREPR}) 
 is that it shows that $P$ is actually sensitive to the value 
 of the subtracted condensates
 $\langle [\bar{\chi} i\gamma_5\tau_3 \chi]_{\rm sub}\rangle$ 
 and $\langle [\bar{\chi} \chi]_{\rm sub} \rangle$. 
 \end{enumerate}
 Before continuing it is useful to pause a moment and discuss the 
 structure of eqs.(\ref{PHIREPR}) and~(\ref{PREPR}) and nature of 
 the various terms appearing in it. 
 \begin{itemize}
 \item We first notice that the contributions from the identity and the
 $F\cdot F$ operator are 
 put together within a square parenthesis in
 eqs.~(\ref{PHIREPR})--(\ref{PREPR}) to remind us that
 there is no unambiguous way to subtract from the latter 
 its power divergent mixing with the identity. Ultimately this 
 is due to the fact that, unlike the chiral condensates, 
 the vacuum expectation value of $F\cdot F$ is not an order 
 parameter of any symmetry. 

 \item For the reason we have just recalled, it is instead perfectly 
 possible to unambigously define, in the massless limit,
 multiplicatively renormalizable operators $[\bar{\chi} \chi]_{\rm sub}$
 and $[\bar{\chi} i\gamma_5 \tau_3 \chi]_{\rm sub}$, by following
 the procedure outlined in refs.~\cite{BMMRT}.
 %\bibitem{BMMRT}  
 %M. Bochicchio, L. Maiani, G. Martinelli, G.C. Rossi  
 %and M. Testa, Nucl. Phys. {\bf B262} (1985) 331;\\
 %M. Testa, JHEP {\bf 9804} (1998) 002.
 More generally, such quark bilinears can be defined as finite 
 operators even at non-vanishing masses, though not uniquely. 
 This can be done by setting, for instance
\begin{eqnarray}
&&[\bar\chi\chi]_{\rm sub}=\bar\chi\chi-
a^{-3} C_{S^0}(r,a\tilde{m},a\mu)\, ,\label{chichi}\\
&&[\bar\chi i\gamma_5 \tau_3 \chi]_{\rm sub} =
\bar\chi i\gamma_5 \tau_3 \chi -
a^{-2}\mu\,C_{P}(r,a\tilde{m},a\mu) \, , \label{chichi5}
\end{eqnarray}
 with the dimensionless coefficient functions $C_{S^0}$ and $C_{P}$
 determined at some finite space-time volume $V=V_0$ by the conditions
\begin{eqnarray}
&&\langle [\bar\chi \chi]_{\rm sub}\rangle_{(r,m_0,\mu)}
= 0 \, ,\qquad V=V_0 \, ,\label{C_S0}\\
&&\langle [\bar\chi i\gamma_5 \tau_3\chi]_{\rm sub}
\rangle_{(r,m_0,\mu)}
= 0 \, ,\qquad V=V_0 \, .\label{C_P3}
\end{eqnarray}
 Both the coefficients $C_{S^0}$ and $C_{P}$ admit a finite
 polynomial expansion in $a\tilde{m}$ and $a\mu$ (actually in
 $(a\mu)^2$ for parity reasons). 

 \item In terms of $[\bar{\chi} \chi]_{\rm sub}$
 and $[\bar{\chi} i\gamma_5 \tau_3 \chi]_{\rm sub}$, the renormalized 
 scalar density in the physical basis, $[\bar\psi \psi]_R$, reads
\begin{equation}
[\bar\psi \psi]_R = Z_M^{-1}(\omega) Z_P
[\cos \omega [\bar\chi \chi]_{\rm sub} +
\sin \omega [\bar\chi i\gamma_5 \tau_3 \chi]_{\rm sub} ] \, ,
\end{equation}
 where $z_m = Z_P/Z_{S^0}$, $Z_M=[z_m^2\cos^2\omega +\sin^2\omega]^{1/2}$
 and $Z_{\Gamma}$ denotes the renormalization constant of 
 $\bar\chi\Gamma\chi$ in the standard Wilson regularization 
 computed in a mass independent renormalization 
 scheme~\footnote{The relations between renormalized and 
 subtracted operators in the $\chi$-basis are 
 $[\bar\chi\chi]_{\rm R}= Z_{S^0}[\bar\chi\chi]_{\rm sub}$ and 
 $[\bar\chi i\gamma_5 \tau_3 \chi]_{\rm R} = 
 Z_P[\bar\chi i\gamma_5 \tau_3 \chi]_{\rm sub}$.}.
 Consistently with the general arguments given above, 
 we remark that only the leading $a^{-3}$ divergent 
 subtraction is uniquely fixed by the symmetries of the theory 
 (WTI's and spurionic transformations). Consequently these 
 properties can be used to make the chiral scalar condensate, 
 $\bar\psi\psi$, multiplicative renormalizable in the massless
 limit, by defining it, e.g.\ as the Wilson average over the 
 expectation values computed with opposite values of the 
 coefficient of the Wilson term \cite{AOKI-PHASE,FREZZOTTI-ROSSI}. 
 \end{itemize} 

 After this little digression let us go back and 
 discuss the implications of eq.~(\ref{PREPR}). 
 If we are on the lattice and take the action of eq.~(\ref{tmaction})
 for values of $\mu$ or $\tilde{m}$ much larger than
 ${\cal O}(a\Lambda_{\rm QCD}^2)$, the condensates
 $\langle [\bar{\chi} i\gamma_5\tau_3 \chi]_{\rm sub} \rangle$
 or $\langle [\bar{\chi} \chi]_{\rm sub} \rangle$ are expected to
 show no metastability and thus the same should be true for the
 plaquette expectation value.
 However, if $\mu$ is smaller than ${\cal O} (a\Lambda^2_{\rm QCD})$ the
 physical scalar condensate signaling spontaneous chiral symmetry
 breaking is not simply given by
 $\langle [\bar\chi i\gamma_5\tau_3\chi]_{\rm sub}\rangle$, but has in
 general also an untwisted component,
 $\langle [\bar\chi\chi]_{\rm sub}\rangle$.
 Both components have an impact on the value of the plaquette (see
 eq.~(\ref{PREPR})).
 When $\tilde{m}$ passes from positive to negative values the
 expectation value of the untwisted operator
 $[\bar{\chi} \chi]_{\rm sub}$ should also change sign and,
 at non-vanishingly small values of $\mu$, eventually
 become very small for almost critical values of $m_0$.
 In this situation, owing to the presence of the chiral symmetry
 breaking term~(\ref{massterms}) in the action, the tmQCD sample
 of gauge configurations is expected to include configurations where
 $\langle [\bar{\chi} \chi]_{\rm sub}\rangle_U$ is positive and
 configurations where $\langle [\bar{\chi} \chi]_{\rm sub}\rangle_U$ is
 negative, corresponding to whether $m_\chi^{\mathrm{PCAC}}$ is positive or
 negative, respectively.
 (By $\langle \ldots \rangle_U$ we mean the fermionic Wick contraction
 on a fixed gauge background $U$.)
 Since the coefficient $b_3=b_3(r,am_0,a\mu)$ does not vanish at
 $m_0 = m _{\mathrm{crit}}$~\footnote{Using the spurionic invariances
 of the action~(\ref{tmaction}), it is possible to show that $b_3$ is
 odd under $(r \to -r) \times (m_0 \to -m_0)$, or equivalently,
 since $m_c(-r) = -m_c(r)$, under
 $(r \to -r) \times (\tilde{m} \to -\tilde{m})$.
 We expect hence a contribution to $b_3$ odd in $r$ and even in
 $\tilde{m}$.}, the value of the plaquette on the configurations where
 $\langle [\bar{\chi} \chi]_{\rm sub}\rangle_U$ is positive should be
 different -- on the basis of the operator expansion~(\ref{PHIREPR}) --
 from that on the configurations where
 $\langle[\bar{\chi} \chi]_{\rm sub}\rangle_U$ is negative.
 The observed jumps of the plaquette expectation value can hence be
 regarded as a combined effect of spontaneous chiral symmetry breaking
 and the explicit breaking of this symmetry due to the Wilson term
 in eq.~(\ref{massterms}).

%%%%%%%%%%%%%%%%%%%%%%%%%%%%%%%%%%%%%%%%%%%%%%%%%%%%%%%%%%%%%%%%%%%%%%%
\subsection{Effective potential model}\label{sec4.2}

 The scenario of a jump in the scalar condensate for Wilson fermions
 on the lattice has actually been given already some time ago by
 Sharpe and Singleton \cite{SHARPE-SINGLETON}.
 As it has been shown in that work, the phase structure of lattice QCD
 for $\mu=0$ with Wilson-type quarks can be understood in the low energy
 chiral theory of pseudo-Goldstone bosons if the influence of leading
 lattice artifacts of ${\cal O}(a)$ and ${\cal O}(a^2)$ is taken into
 account.

 There are two alternatives: either there exists an {\em Aoki phase}
 \cite{AOKI-PHASE} or there is a first order phase transition between
 the phases with positive and negative quark mass and the Aoki phase
 does not exist.

 The relevant part of the effective potential is written in
 \cite{SHARPE-SINGLETON} as
\be\label{eq4:1}
{\cal V}_\chi = -c_1 A + c_2 A^2 \ .
\ee
 Here $A$ denotes the flavour singlet component of the SU(2) matrix
 valued field $\Sigma$ in the low energy effective chiral Lagrangian:
\be\label{eq4:2}
\Sigma = A + i \sum_{r=1}^3 B_r \tau_r \ .
\ee
 Because of the relation $1 = A^2 + \sum_{r=1}^3 B_r B_r$ the variable
 $A$ is constrained to lie between -1 and +1 inclusive.
 In the vicinity of the critical quark mass the constant
 $c_2 = {\cal O}(a^2)$ and the other parameter $c_1$ is proportional to
 the bare quark mass (in our notations $c_1 \propto \tilde{m}$).

 In order to find the ground state (``vacuum'') the effective potential
 has to be minimized.
 Without repeating the details of the discussion in
 ref.~\cite{SHARPE-SINGLETON} let us just summarize the result.

 In case of positive $c_2$ there exists an Aoki phase in the
 region of bare quark masses defined by $-2c_2 \leq c_1 \leq 2c_2$.
 At the boundaries $c_1=\pm 2c_2$ all three pion masses vanish.
 Inside the Aoki phase the charged pions are massless because they are
 the Goldstone bosons of spontaneous flavour symmetry breaking but the
 neutral pion is massive.
 Outside the Aoki phase ($|c_1| > 2c_2$) the flavour symmetry is
 preserved by the ground state and the three degenerate pions are
 massive (see figure \ref{phases}).

 The other alternative is that $c_2$ is negative.
 In this case the flavour symmetry is preserved everywhere but
 there exists a minimal pion mass because the pion mass is given by
\be\label{eq4:3}
m_\pi^2 = f_\pi^{-2} \left( |c_1| + 2|c_2| \right) \ .
\ee

 At $c_1 = 0$ the vacuum expectation value jumps from
 $\Sigma=A=+1$ to $\Sigma=A=-1$.
 Since the jump of this ``order parameter'' happens at non-zero pion
 mass (i.e.~finite correlation length) the thermodynamical description
 of the behaviour near $c_1=0$ corresponds to a {\em first order phase
 transition}.

 An interesting intermediate situation is defined by $c_2=0$.
 In this case the vacuum expectation value jumps between
 $\Sigma=A=+1$ and $\Sigma=A=-1$ at a single first order phase
 transition point.
 This limiting case is the ideal situation, when the phase structure
 in the Sharpe-Singleton model is identical to the expected one in the
 continuum.
 It can be characterized either by saying that the
 Aoki phase has zero extension or that the minimal pion mass is
 zero (see figure \ref{phases}).
 Of course, this behaviour is valid only up to ${\cal O}(a^3)$ effects,
 neglecting higher orders in the chiral expansion.

%%%%%%%%%%%%%%%%%%%%%%%%%%%%%%%%%%%%%%%%%%%%%%%%%%%%%%%%%%%%%%%%%%%%%%%%
\subsection{Scenarios}\label{sec4.3}

 Our numerical results reveal that we clearly observe metastabilities
 in various quantities.
 Thus our conclusion is that at least for vanishing twisted mass
 parameter, i.e.~for the standard Wilson lattice theory, {\em there is a
 first order phase transition.}
 For non-vanishing values of $\mu$ we can have two scenarios.

 The first is that the first order phase transition persists for
 $\mu \ne 0$ but sufficiently small in absolute value.
 For large $\mu$ the theory approaches the quenched limit with a
 constant quark determinant and therefore it is plausible that no phase
 transition is expected.
 This scenario suggests that the first order phase transition line in
 the $(m_0,\mu)$ plane has an end point: the two phases with positive
 and negative quark masses are analytically connected
 (see figure \ref{phases}).
 The situation is in this sense analogous to the phase structure of
 the SU(2) fundamental Higgs model (see chapter 6 of
 \cite{MONTVAY-MUNSTER} and references therein).

 The second scenario is that for any non-vanishing value of $\mu$ the
 first order phase transition disappears.
 In this scenario, when varying $m_0$, one passes at some small
 distance from the first order phase transition at $\mu=0$ and just
 feels this close-by phase transition.

 We can at present not differentiate between these two scenarios.
 From the numerical side we would need to know better the $\mu$ and
 $\beta$ dependence of the metastability phenomena.
 From the analytical side an analysis \`a la Sharpe and Singleton
 including the twisted mass parameter $\mu$ is helpful.\footnote{
 We thank Gernot M\"unster for discussions on this and for communicating
 us his results before publication.}

%%%%%%%%%%%%%%%%%%%%%%%%%%%%%%%%%%%%%%%%%%%%%%%%%%%%%%%%%%%%%%%%%%%%%%%%
\begin{figure}
\vspace*{-30mm}
\begin{center}
\epsfig{file=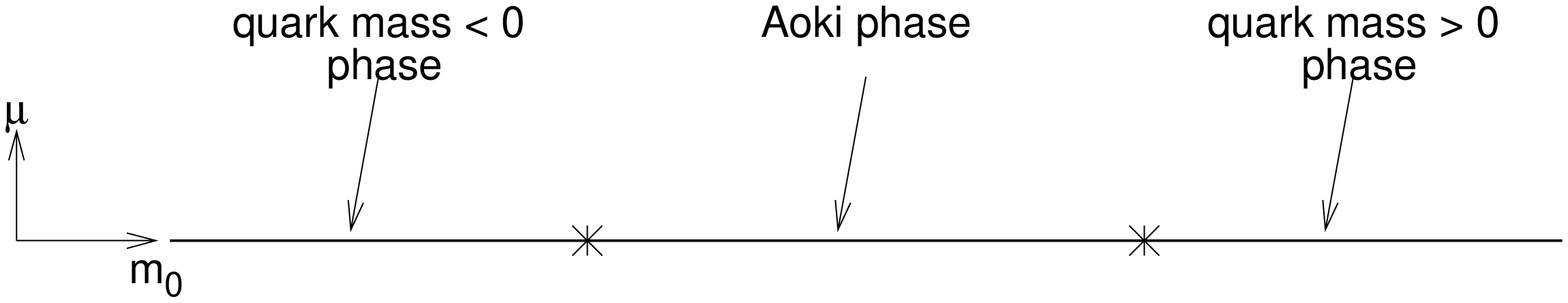,
        width=90mm,height=120mm,angle=-90}
        %bbllx=50pt,bblly=-60pt,bburx=554pt,bbury=690pt}
\end{center}
\vspace*{-50mm}
\begin{center}
\epsfig{file=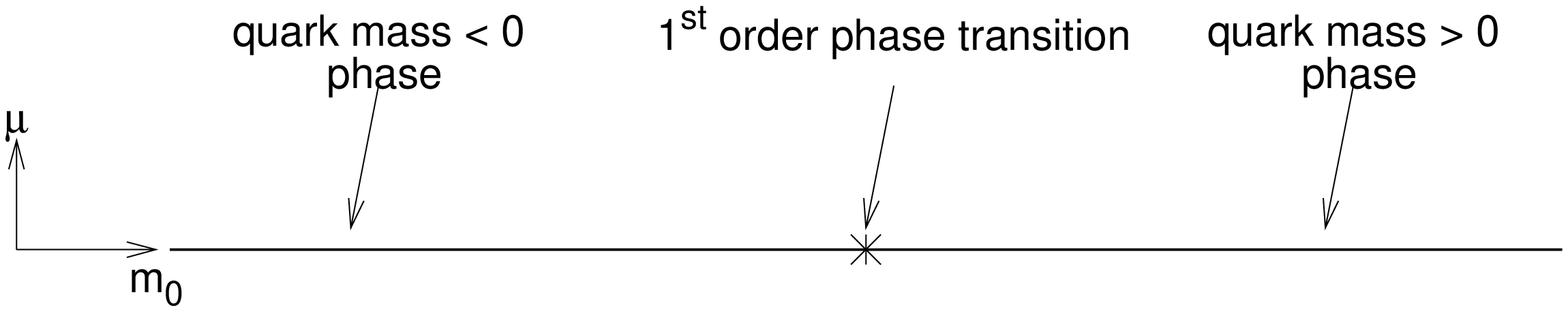,
        width=90mm,height=120mm,angle=-90}
        %bbllx=50pt,bblly=-60pt,bburx=554pt,bbury=690pt}
\end{center}
\vspace*{-45mm}
\begin{center}
\epsfig{file=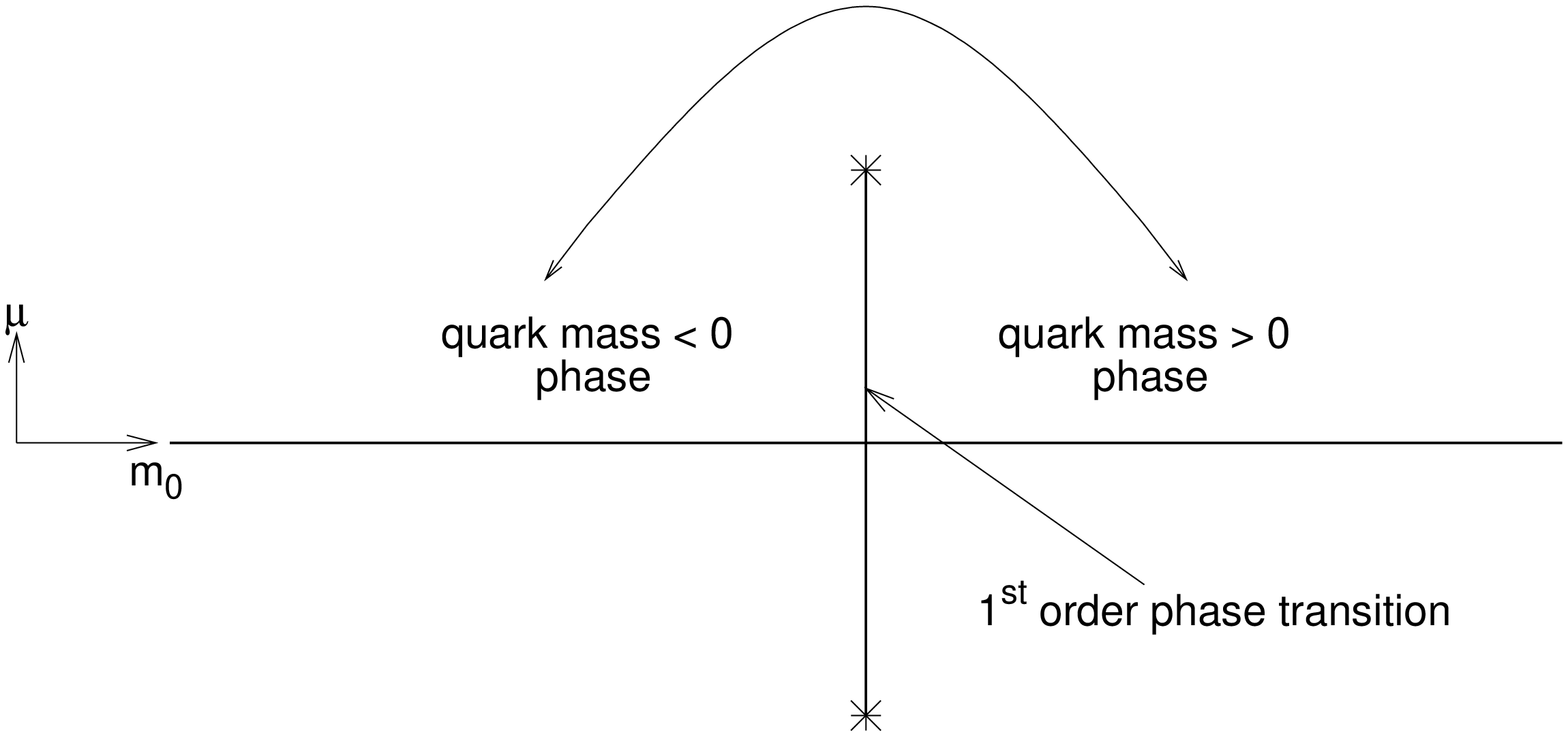,
        width=90mm,height=120mm,angle=-90}
        %bbllx=50pt,bblly=-60pt,bburx=554pt,bbury=690pt}
\end{center}
%
%\vspace*{-30mm}
\vspace*{-15mm}
\begin{center}
\parbox{12cm}{\caption{
 The alternatives of the phase structure in the $(m_0,\mu)$-plane:
 {\em Upper part:} Aoki phase at $\mu=0$ if $c_2 > 0$,
 {\em middle part:} first order phase transition point if $c_2=0$,
 {\em lower part:} first order phase transition line if $c_2 < 0$.
 In the latter case the two phases are connected with each other as it
 is shown by the curve with arrows at both ends.
\label{phases}}}
\end{center}
\end{figure}
%%%%%%%%%%%%%%%%%%%%%%%%%%%%%%%%%%%%%%%%%%%%%%%%%%%%%%%%%%%%%%%%%%%%%%%%

 The first order phase transition between the phases with positive
 and negative quark masses observed in the previous section is
 consistent with the {\em no-Aoki-phase} alternative ($c_2 < 0$) of
 Sharpe and Singleton.

 Our exclusion of the Aoki phase is in agreement with the results of
 a recent paper \cite{HUMBOLDT} which suggests that in case of the
 unimproved Wilson action the Aoki phase is restricted to the region
 of strong gauge couplings ($\beta \leq 4.6$).
 Note that in an early paper on QCD thermodynamics with Wilson quarks
 \cite{BLUMETAL} a first order ``bulk'' phase transition has also been
 observed at $\beta = 4.8$ which is consistent both with
 ref.~\cite{HUMBOLDT} and with our observations.
 For further numerical work on the Aoki phase, see refs.~\cite{aokiwork}

 The rather strong metastability of the two phases with positive and
 negative quark mass can be understood on the basis of the
 properties of the eigenvalue spectrum of the (non-hermitean)
 Wilson-fermion matrix in the twisted mass basis corresponding to
 eq.~(\ref{tmaction}).
 For zero twisted mass ($\mu=0$) at small positive quark masses
 there is a ``cloud'' of eigenvalues close to the origin near the real
 axis.
 (For a numerical study see section 4 of \cite{QQ+Q1}.)
 In order to reach negative quark masses this ``cloud'' has to cross
 near the origin to the other side with negative real parts.
 This {\em eigenvalue cloud crossing} is strongly suppressed by the
 zero of the determinant.
 This, we believe, is the reason at the microscopical level for the
 observed strong metastability.
 For non-zero twisted mass there is a strip of width $2|\mu|$ around
 the real axis where there are no eigenvalues.
 If this strip is wide enough the eigenvalues are sufficiently far away
 from the origin and the first order phase transition disappears.

 As it was already emphasized in \cite{SHARPE-SINGLETON}, the sign of
 the coefficient $c_2$ in the low energy pion effective potential is
 not universal, it depends on the way the action is discretized.
 Therefore a clever choice of the lattice action may weaken the first
 order phase transition and, for instance, decrease the minimal pion
 mass at it.
 Previous results of the JLQCD Collaboration \cite{JLQCD} support the
 conjecture that changing the gauge action alone has an important
 effect.
 If, indeed, one could find some parameter in the lattice gauge action
 which at some value would change the sign of $c_2$ an appealing
 possibility would be to tune the lattice action to this value.
 The features of a discretization with $c_2=0$ seem to be quite
 favourable from the point of view of light quark simulations when, up
 to ${\cal O}(a^2)$, there would be just a single point in the
 $(m_0,\mu)$ plane with vanishing pion mass -- an ideal situation
 corresponding to the expected phase structure in the continuum.

%%%%%%%%%%%%%%%%%%%%%%%%%%%%%%%%%%%%%%%%%%%%%%%%%%%%%%%%%%%%%%%%%%%%%%%%
\section{Conclusion}\label{sec5}

 In this paper we have explored Wilson twisted mass fermions restricting
 ourselves to simulations at only one value of $\beta=5.2$.
 By fixing the twisted mass parameter $\mu$ and changing the untwisted
 Wilson quark mass $m_0$, or equivalently the hopping parameter
 $\kappa$, we encountered strong metastabilities in the plaquette
 expectation value, visible both in thermal cycles as well as in
 long-living metastable states.
 At the same time, the pion mass does not vanish but has a minimum at
 a rather large value.
 The PCAC quark mass $m_\chi^{\mathrm{PCAC}}$ in the different metastable
 branches is positive for the branch with low plaquette expectation
 value and it is negative for the branch with high plaquette
 expectation value.

 The detection of these metastabilities became possible by employing a
 twisted mass term.
 Only a non-vanishing value of $\mu$ allowed us to cross the critical
 quark mass.
 We showed that for lattice theories that break chiral symmetry
 explicitly the jump of the scalar condensate, when changing the sign
 of the quark mass, induces a jump of the plaquette expectation
 value with associated signs of metastability.
 For $\mu=0$ these metastabilities find a natural interpretation in the
 effective potential model of Sharpe and Singleton, arising from
 spontaneous symmetry breaking and using a low energy effective
 Lagrangian which also describes lattice artifacts.
 The agreement with the Sharpe-Singleton model is remarkable because
 in the continuum limit in this model the phase structure of lattice
 QCD with Wilson quarks approaches fast -- at a rate ${\cal O}(a^2)$ --
 the expected phase structure of QCD near zero quark mass.
 This is an important property which has to be required from any
 lattice regularization of QCD.

 It should be clear that our work can only represent a first step in a
 detailed understanding of the QCD phase diagram at zero temperature
 near vanishing quark masses.
 Clearly, substantially more work has to be done to resolve this phase
 structure and its behaviour in the continuum limit.
 For instance, at present for $\mu \ne 0$ we are unable to differentiate
 between a scenario where the first order phase transition persists and
 another one where at $\mu \ne 0$ only a remnant of the phase transition
 at $\mu=0$ is seen.
 In this respect an analysis like in ref.~\cite{SHARPE-SINGLETON} for
 $\mu \ne 0$ is very helpful \cite{MUNSTER}.

 Among the many open questions there are:
 How fast does the gap vanish when the continuum limit at higher values
 of $\beta$ is approached?
 How are the signs of metastability related to the ones observed using
 the Wilson plaquette action and clover-improved Wilson fermions?
 How precisely do the eigenvalues re-arrange when the critical quark
 mass is crossed?
 Do different gauge actions change the couplings of the effective
 potential and may hence lead to avoid the phenomena of metastability
 and reproduce the ideal phase structure at vanishing quark mass
 already for non-zero lattice spacing?

 The most important question is, of course, how phenomenology can be
 done given the metastability phenomenon seen in our present results;
 i.e.~what is the lowest value of the quark mass that can be reached
 before one enters the regime of metastabilities and how does this
 change with decreasing value of the lattice spacing.

%%%%%%%%%%%%%%%%%%%%%%%%%%%%%%%%%%%%%%%%%%%%%%%%%%%%%%%%%%%%%%%%%%%%%%%%
\vspace*{1em}\noindent
{\large\bf Acknowledgments}

\noindent
 We thank Gernot M\"unster, Luigi Scorzato and Stephen Sharpe for helpful
 discussions.
 The computations were performed on the APEmille systems installed 
 at NIC Zeuthen and INFN Milano, the IBM-JUMP computer at NIC
 J\"ulich, IBM pSeries 690 Supercomputer at HLRN, the PC clusters at
 DESY Hamburg, NIC Zeuthen, University of M\"unster,
 Forschungszent\-rum Karlsruhe and the Sun Fire SMP-Cluster at the
 Rechenzentrum - RWTH Aachen.
 This work was supported by the DFG Sonderforschungsbereich/Transregio
 SFB/TR9-03.
G.C.R. wishes to
thank the Humboldt Foundation for financial support while this work was
prepared.

\newpage
%%%%%%%%%%%%%%%%%%%%%%%%%%%%%%%%%%%%%%%%%%%%%%%%%%%%%%%%%%%%%%%%%%%%%%%%

%%%%%%%%%%%%%%%%%%%%%%%%%%%%%%%%%%%%%%%%%%%%%%%%%%%%%%%%%%%%%%%%%%%%%%%%

\newpage
\appendix
%%%%%%%%%%%%%%%%%%%%%%%%%%%%%%%%%%%%%%%%%%%%%%%%%%%%%%%%%%%%%%%%%%%%%%%%
\section{Appendix}
\subsection{Even-odd preconditioning for the HMC algorithm}
\label{app.hmc}

 Let us start with the Dirac operator in the hopping parameter
 representation in the twisted basis written as
\begin{equation}\label{eq:eo0}
S[\chi,\bar\chi,U] \equiv 
\sum_{xy} \bar{\chi}(x)\, M_{xy}\, \chi(y)
\end{equation}
 where the matrix $M$ can be easily read from eq. (\ref{eq2.1:6}).
 Using $M$ one can define the hermitian operator
\begin{equation}
  \label{eq:eo1}
  Q\equiv \gamma_5 M = \begin{pmatrix}
      Q_+ & 0   \\
      0   & Q_- \\
      \end{pmatrix}
\end{equation}
 where the submatrices $\Qpm$ can be factorised as follows:
\begin{equation}
  \label{eq:eo2}
  \begin{split}
    Q_\pm &= \gamma_5\begin{pmatrix}
      1\pm i\tilde\mu\gamma_5 & M_{eo} \\
      M_{oe}    & 1\pm i\tilde\mu\gamma_5 \\
    \end{pmatrix} =
    \gamma_5\begin{pmatrix}
      M_{ee}^\pm & M_{eo} \\
      M_{oe}    & M_{oo}^\pm \\
    \end{pmatrix} \\[0.5em]
    & =
    \begin{pmatrix}
      \gamma_5M_{ee}^\pm & 0 \\
      \gamma_5M_{oe}  & 1 \\
    \end{pmatrix}
    \begin{pmatrix}
      1       & (M_{ee}^\pm)^{-1}M_{eo}\\
      0       & \gamma_5(M_{oo}^\pm-M_{oe}(M_{ee}^\pm)^{-1}M_{eo})\\
    \end{pmatrix}
\end{split}
\end{equation}
 and we have defined $\tilde\mu \equiv 2\kappa\mu$.
 Note that $(M_{ee}^\pm)^{-1}$ can be easily computed to 
\[
(1\pm i\tilde\mu\gamma_5)^{-1} = \frac{1\mp i\tilde\mu\gamma_5}{1+\tilde\mu^2}.
\]
 Using $\det(Q) = \det(Q_+)\det(Q_-)$ one can now derive the following
 relation (an equation apart from an irrelevant factor):
\begin{equation}
  \label{eq:eo4}
  \begin{split}
\det(Q_\pm) &\propto \det(\hat Q_\pm) \\[0.5em]
\hat Q_\pm &:= \gamma_5(M_{oo}^\pm - M_{oe}(M_{ee}^\pm )^{-1}M_{eo})\, ,
  \end{split}
\end{equation}
 where $\hat Q_\pm$ is only defined on the odd sites of the lattice.
 In the HMC algorithm the determinant is stochastically estimated using
 pseudo-fermion fields $\phi_o$:
\[
\begin{split}
  \det(\hat Q_+ \hat Q_-) &= \int D[\phi_o, \phi_o^\dagger]\exp(-S_b)\, ,\\
  S_b &:=\phi_o^\dagger (\hat Q_+\hat Q_-)^{-1} \phi_o\, ,
\end{split}
\]
 where the fields $\phi_o$ are defined only on the odd sites of the
 lattice.
 In order to compute the force corresponding to the effective action
 $S_b$ we need the variation of $S_b$ with respect to the gauge fields
 (using $\delta (A^{-1})=-A^{-1}\delta A A^{-1}$):
\begin{equation}
  \label{eq:eo5}
  \begin{split}
    \delta S_b & = -[\phi_o^\dagger (\hat Q_+ \hat Q_-)^{-1}
                  \delta \hat Q_+ \hat Q_+^{-1}\phi_o +
    \phi_o^\dagger\hat Q_-^{-1}\delta \hat Q_-
     (\hat Q_+ \hat Q_-)^{-1} \phi_o ] \\[0.5em]
     & = -[X_o^\dagger \delta \hat Q_+ Y_o + Y_o^\dagger 
       \delta\hat Q_- X_o]
  \end{split}
\end{equation}
 with $X_o$ and $Y_o$ defined on the odd sites as
\begin{equation}
  \label{eq:eo6}
  X_o = (\hat Q_+ \hat Q_-)^{-1} \phi_o,\quad Y_o =
        \hat Q_+^{-1}\phi_o=\hat Q_-X_o\ ,
\end{equation}
 where $\hat Q_\pm^\dagger = \hat Q_\mp$ has been used.
 The variation of $\hat Q_\pm$ reads
\begin{equation}
  \label{eq:eo7}
  \delta \hat Q_\pm = 
    \gamma_5\left(-\delta M_{oe}(M_{ee}^\pm )^{-1}M_{eo} -
    M_{oe}(M_{ee}^\pm )^{-1}\delta M_{eo}\right),
\end{equation}
 and one finds
\begin{equation}
  \label{eq:eo8}
  \delta S_b = -(X^\dagger\delta Q_+ Y + Y^\dagger\delta Q_- X) =
-(X^\dagger\delta Q_+ Y +(X^\dagger\delta Q_+ Y)^\dagger)
\end{equation}
 where $X,Y$ is now defined over the full lattice as
\begin{equation}
  \label{eq:eo9}
  X = 
  \begin{pmatrix}
    -(M_{ee}^-)^{-1}M_{eo}X_o \\ X_o\\
  \end{pmatrix},\quad
  Y = 
  \begin{pmatrix}
    -(M_{ee}^+)^{-1}M_{eo}Y_o \\ Y_o\\
  \end{pmatrix}.
\end{equation}
 In addition,
 $\delta Q_+ = \delta Q_-$, $M_{eo}^\dagger = \gamma_5 M_{oe}\gamma_5$
 and $\quad M_{oe}^\dagger = \gamma_5M_{eo}\gamma_5$ has been used.
 Since the bosonic part is quadratic in the $\phi_o$ fields, the
 $\phi_o$ are generated at the beginning of each molecular dynamics
 trajectory with
\begin{equation}
  \label{eq:eo10}
  \phi_o = \hat Q_+ R,
\end{equation}
 where $R$ is a random spinor field taken from a Gaussian distribution
 with norm one.

%%%%%%%%%%%%%%%%%%%%%%%%%%%%%%%%%%%%%%%%%%%%%%%%%%%%%%%%%%%%%%%%%%%%%%%%
\subsubsection{Hasenbusch trick}

 The trick first presented in \cite{HASENBUSCH} is based on the
 observation that writing
\begin{equation}
  \label{eq:mt1}
  \det[\Qp\Qm] = \det[\Wp\Wm]\cdot\det[(\Qp\Qm)/(\Wp\Wm)]
\end{equation}
 is advantageous for the HMC, if the condition number of $\Wp\Wm$ and of
 $(\Qp\Qm)/(\Wp\Wm)$ is significantly reduced compared to the condition
 number of only $(\Qp\Qm)$.
 In order to achieve this we define
\begin{equation}
  \label{eq:mt2}
  \begin{split}
    \Qpm &= \gamma_5 D_W \pm i\tilde\mu\, ,\\
    \Wpm &= \gamma_5 D_W \pm i\tilde\mu_2\, .\\
  \end{split}
\end{equation}
 With $\tilde\mu_2 =\tilde\mu+\Delta\tilde\mu$ it follows immediately that the condition
 number of $\Wp\Wm$ is lower than the one of $\Qp\Qm$ if for
 $\lambda_{\textrm{min}}$ and $\lambda_{\textrm{max}}$ the lowest and
 the largest eigenvalue of $\Qp\Qm$, respectively,
 $|\lambda_{\textrm{min}}|\ll\tilde\mu_2^2\ll|\lambda_{\textrm{max}}|$ holds:
 the condition number of $\Wp\Wm$ is $|\lambda_{\textrm{max}}|/\tilde\mu_2^2$
 while the one of
 $(W_+W_-)^{-1} (Q_+Q_-)^{2}$ 
 contrariwise is $\tilde\mu_2^2/|\lambda_{\textrm{min}}|$.
 We can take $\tilde\mu$ which is a lower bound for $|\lambda_{\textrm{min}}|$
 to write down the following estimates for the condition numbers $k$:
\[
k_{\Wp\Wm} = \frac{|\lambda_{\textrm{max}}|}{\tilde\mu_2^2}\, ,\quad
k_{(\Qp\Qm)/(\Wp\Wm)} \leq \frac{\tilde\mu_2^2}{\tilde\mu^2}\, ,
\]
 which leads to an optimal choice for
 $\tilde\mu_2^2=\sqrt{|\lambda_{\textrm{max}}|\cdot\tilde\mu^2}$.
 As has been shown in \cite{HASENBUSCH-JANSEN} also the force
 contribution coming from $(\Qp\Qm)/(\Wp\Wm)$ is reduced.
 This is true also for tmQCD and can bee seen in the following
 way: noticing that
\begin{equation}
  \label{eq:mt3}
  \begin{split}
    \Qp\Qm &= Q^2+\tilde\mu^2\, \qquad\textrm{and}\\ 
    \Wp\Wm &= Q^2+\tilde\mu_2^2 = Q^2+\tilde\mu^2+\tilde\mu_2^2-\tilde\mu^2
    = \Qp\Qm +\tilde\mu_2^2-\tilde\mu^2\, , 
  \end{split}
\end{equation}
 it follows that
\begin{equation}
  \label{eq:mt4}
  \Wp\Wm(\Qp\Qm)^{-1}= 1 + (\tilde\mu_2^2 - \tilde\mu^2)(\Qp\Qm)^{-1}.
\end{equation}
 Since the corresponding effective action reads
\begin{equation}
  \label{eq:mt5}
  S_F = \phi^\dagger( 1 + (\tilde\mu_2^2 - \tilde\mu^2)(\Qp\Qm)^{-1})\phi 
\end{equation}
 one can see that one gets an explicit factor
 $(\tilde\mu_2^2 - \tilde\mu^2)\ll1$ multiplying the force contribution
 compared to the original effective action which will reduce the force
 and therefore lead to a smoother evolution of the algorithm.

 Let us remark that the procedure explained above can be immediately
 applied to the even-odd preconditioned system.
 Furthermore the trick can be iterated to two or even more additional
 operators. 

%%%%%%%%%%%%%%%%%%%%%%%%%%%%%%%%%%%%%%%%%%%%%%%%%%%%%%%%%%%%%%%%%%%%%%%%
\begin{figure}[htbp]
  \centering
  \includegraphics[]{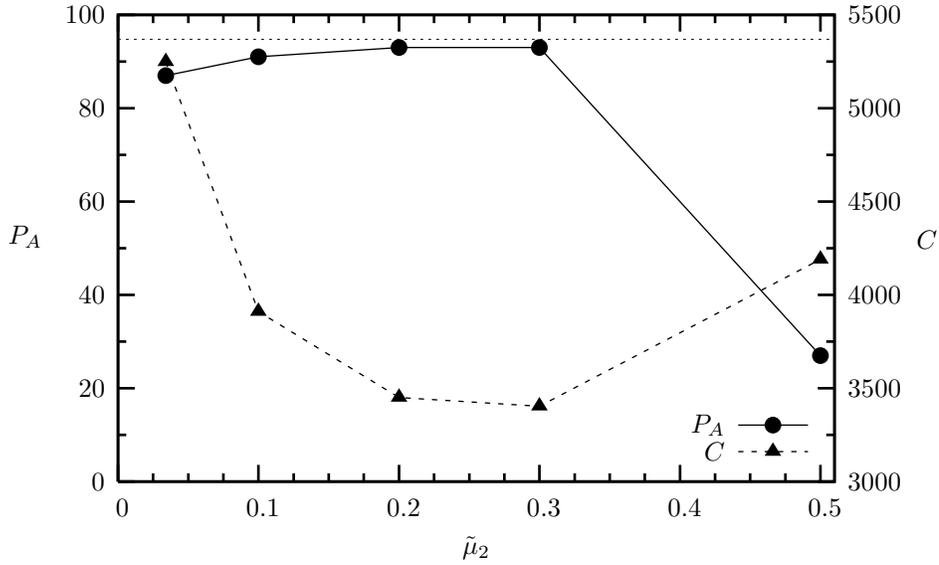}
  \caption{Acceptance rate $P_A$ and cost $C$ in units of CG
    iterations versus $\tilde\mu_2 = 2\kappa\mu_2$ at fixed HMC stepsize
    and trajectory length.
    The dashed line represents the cost required to obtain about 90\%
    acceptance rate without the additional operator.
    The parameters are: $8^4$ lattice, $\beta=5.2$, $\kappa=0.17$,
    $\mu=0.01$.}
  \label{fig:martinstrick}
\end{figure}
%%%%%%%%%%%%%%%%%%%%%%%%%%%%%%%%%%%%%%%%%%%%%%%%%%%%%%%%%%%%%%%%%%%%%%%%

 In fig.~\ref{fig:martinstrick} the cost $C$ in units of CG iterations
 and the acceptance rate $P_A$ is plotted versus
 $\tilde\mu_2 = 2\kappa\mu_2$ at fixed HMC stepsize and trajectory
 length.
 One can see that as expected the acceptance rate increases by
 introducing an additional operator and reaches a maximum around
 $\tilde\mu=0.2$.
 Of course also the costs increase when compared to the HMC without
 additional operators.
 But the costs are still much less than what is needed to reach an
 acceptance rate of about $90$\% without the additional operator
 (see the dashed line 
in fig.~\ref{fig:martinstrick}). One can see that the gain in the costs is
 about a factor of two.

%%%%%%%%%%%%%%%%%%%%%%%%%%%%%%%%%%%%%%%%%%%%%%%%%%%%%%%%%%%%%%%%%%%%%%%%
\subsection{Even-odd preconditioning for the TSMB algorithm}
\label{app.tsmb}

 In this appendix even-odd preconditioning is derived for the TSMB
 algorithm.
 The even-odd subspace decomposition of the fermion matrix in the
 twisted basis can be written as
\be\label{eqa.2:1}
Q^\chi = \left( \begin{array}{cc}
\mu_1+i\gamma_5\tau_3\mu  &  -\half M_{eo} \\
-\half M_{oe}  &  \mu_1+i\gamma_5\tau_3\mu
\end{array} \right)
\ee
 where indices start by zero = even, the lattice spacing is set to
 $a=1$ and the abbreviation $\mu_1 \equiv m_0+4r = (2\kappa)^{-1}$ is
 introduced.
 The {\em hermitean} fermion matrix
 $\tilde{Q}=\gamma_5\tau_1 Q^\chi=\tilde{Q}^\dagger$
 is then
\be\label{eqa.2:2}
\tilde{Q} = \left( \begin{array}{cc}
\gamma_5\tau_1\mu_1+\tau_2\mu  &  -\half\gamma_5\tau_1 M_{eo} \\
-\half\gamma_5\tau_1 M_{oe}  &  \gamma_5\tau_1\mu_1+\tau_2\mu
\end{array} \right)  \ .
\ee
 Using the notation
\be\label{eqa.2:3}
t_5 \equiv (\gamma_5\tau_1\mu_1+\tau_2\mu)^{-1} \gamma_5\tau_1
= (\mu_1-i\gamma_5\tau_3\mu)(\mu_1^2+\mu^2)^{-1}
\ee
 one can write $\tilde{Q}$ as the following product:
\begin{eqnarray}
\tilde{Q} & = &
\left( \begin{array}{cc}
\gamma_5\tau_1\mu_1+\tau_2\mu  &  0  \\
0  &  \gamma_5\tau_1\mu_1+\tau_2\mu
\end{array} \right)
\left( \begin{array}{cc}
1  &  0                 \\
-\half t_5 M_{oe}  &  1
\end{array} \right)
\nonumber \\[1.0em]\label{eqa.2:4}
& \cdot &
\left( \begin{array}{cc}
1  &  0                                 \\
0  &  1-\frac{1}{4}t_5 M_{oe}t_5 M_{eo}
\end{array} \right)
\left( \begin{array}{cc}
1  &  -\half t_5 M_{eo} \\
0  &  1
\end{array} \right)  \ .
\end{eqnarray}

 This can be used for preconditioned inversion of $\tilde{Q}$
 because the inverse of all the factors but the third one is
 trivial.
 Of course, the third factor is expected to have smaller condition
 number than $\tilde{Q}$ itself.

 Multi-boson (MB) updating can be set up following \cite{JEGERLEHNER}.
 Since the determinant of the above triangular matrices is equal
 to 1 we have
\begin{eqnarray}
\det\tilde{Q} & = & \det\left( \begin{array}{cc}
\gamma_5\tau_1\mu_1+\tau_2\mu & 0 \\
0 & \gamma_5\tau_1\mu_1+\tau_2\mu-\frac{1}{4}\gamma_5\tau_1 M_{oe}
(\gamma_5\tau_1\mu_1+\tau_2\mu)^{-1}\gamma_5\tau_1 M_{eo}
\end{array} \right)
\nonumber \\[2.0em]
& = & \det_e\left(\gamma_5\tau_1\mu_1+\tau_2\mu\right)
\nonumber \\[0.5em]\label{eqa.2:5}
& \cdot & \det_o\left( \gamma_5\tau_1\mu_1+\tau_2\mu
-\frac{1}{4}\gamma_5\tau_1 M_{oe}
(\gamma_5\tau_1\mu_1+\tau_2\mu)^{-1}\gamma_5\tau_1 M_{eo}
\right)
\end{eqnarray}
 where $\det_e$ and $\det_o$ denote determinants in the even and
 odd subspaces, respectively.
 The first factor does not depend on the gauge field and therefore
 it can be omitted.
 In the second factor we have the hermitean matrix defined on odd
 sites
\begin{eqnarray}\label{eqa.2:6}
\bar{Q} &=& \gamma_5\tau_1\mu_1+\tau_2\mu
-\frac{1}{4}\gamma_5\tau_1 M_{oe}
(\gamma_5\tau_1\mu_1+\tau_2\mu)^{-1}\gamma_5\tau_1 M_{eo} = \nonumber \\
&=& \gamma_5\tau_1\mu_1+\tau_2\mu
-\frac{1}{4}\gamma_5\tau_1 M_{oe}
(\gamma_5\tau_1\mu_1+\tau_2\mu)(\mu_1^2+\mu^2)^{-1}
\gamma_5\tau_1 M_{eo} = \bar{Q}^\dagger \ .
\end{eqnarray}
 The hermiticity of $\bar{Q}$, which can be called
 {\em hermitean preconditioned fermion matrix}, follows from
\be\label{eqa.2:7}
\gamma_5\tau_1 M_{oe}^\dagger \gamma_5\tau_1 = M_{eo} \ .
\ee

 In MB updating one can start with the identity
\be\label{eqa.2:8}
\det Q = \det\tilde{Q} \,\propto\, \det_o\bar{Q} =
\left(\det_o\bar{Q}^2\right)^\half \simeq
\frac{1}{\det_o P_\half (\bar{Q}^2) }
\ee
 where the $P_\half$ is a polynomial approximation satisfying
\be\label{eqa.2:9}
P_\half(x) \simeq \frac{1}{x^\half}
\ee
 in an interval $x \in [\epsilon,\lambda]$ covering the spectrum
 of $\bar{Q}^2$.
 (Note that for $\mu \ne 0$ $\det Q$ and $\det\bar{Q}$
 are positive.)

 The rest is the same as usual: one writes the polynomial with the help
 of the square roots of its roots $\rho_j,\; j=1,2,\ldots$ as
\be\label{eqa.2:10}
P_\half(\bar{Q}^2) \,\propto\,
\prod_j (\bar{Q}-\rho_j^*)(\bar{Q}-\rho_j) \ .
\ee
 Then using the identity
\be\label{eqa.2:11}
\det\left( \begin{array}{cc}
A_{ee}  & A_{eo}  \\
A_{oe}  & A_{oo}
\end{array} \right) =
\det_e A_{ee} \cdot
\det_o\left( A_{oo}-A_{oe}A_{ee}^{-1}A_{eo} \right)
\ee
 one obtains
\begin{eqnarray}
\det_o(\bar{Q}-\rho_j) & = &
\det_e\left(\gamma_5\tau_1\mu_1+\tau_2\mu\right)^{-1}
\nonumber \\[1.0em]\label{eqa.2:12}
& \cdot & \det\left( \begin{array}{cc}
\gamma_5\tau_1\mu_1+\tau_2\mu  &  -\half\gamma_5\tau_1 M_{eo} \\
-\half\gamma_5\tau_1 M_{oe}  &  \gamma_5\tau_1\mu_1+\tau_2\mu-\rho_j
\end{array} \right) \ .
\end{eqnarray}
 Denoting the projector on the odd subspace by $P_o$ we finally
 obtain the multi-boson representation
\begin{eqnarray}
\left(\det_o\bar{Q}^2\right)^\half & \propto &
\prod_j \frac{1}{\det\left[(\tilde{Q}-P_o\rho_j^*)
(\tilde{Q}-P_o\rho_j)\right]}
\nonumber \\[1.0em]\label{eqa.2:13}
& \propto & \int [d\Phi] \exp \left\{
-\sum_j \Phi_j^\dagger (\tilde{Q}-P_o\rho_j^*)
(\tilde{Q}-P_o\rho_j)\Phi_j \right\} \ .
\end{eqnarray}
%
%%%%%%%%%%%%%%%%%%%%%%%%%%%%%%%%%%%%%%%%%%%%%%%%%%%%%%%%%%%%%%%%%%%%%%%%

\end{document}